# Multivariate MR Biomarkers Better Predict Cognitive Dysfunction in Mouse Models of Alzheimer's Disease

Short title:

MRI Predictors of Cognitive Dysfunction in Mouse Models of Alzheimer's Disease

Alexandra Badea[1, 2, 3] *, Natalie A Delpratt[1, *], RJ Anderson[1], Russell Dibb[1], Yi Qi[1], Hongjiang Wei[4], Chunlei Liu[5], William C Wetsel[6], Brian B Avants[7], Carol Colton[2]

[1]Center for In Vivo Microscopy, Department of Radiology, Duke University Medical Center, Durham NC

[2] Department of Neurology, Duke University Medical Center, Durham NC

[3] Brain Imaging and Analysis Center, Duke University, Durham, NC.

[4]Institute for Medical Imaging Technology, School of Biomedical Engineering, Shanghai Jiao Tong University, Shanghai, China (hongjiang.wei@sjtu.edu.cn)

[5]Department of Electrical Engineering and Computer Science, University of California, Berkeley, CA

[6]Department of Psychiatry and Behavioral Sciences, Cell Biology, Neurobiology, Duke University Medical Center, Durham, NC

[7] Department of Radiology and Medical Imaging, University of Virginia, Charlottesville, VA

* The first two authors share first authorship

Corresponding Author:

Alexandra Badea, PhD

Center for In Vivo Microscopy

Department of Radiology

BOX 3302 Duke University Medical Center

Durham, NC 27710

USA

alexandra.badea@duke.edu

**Abstract**

To understand multifactorial conditions such as Alzheimer's disease (AD) we need brain signatures that predict the impact of multiple pathologies and their interactions. To help uncover the relationships between pathology affected brain circuits and cognitive markers we have used mouse models that represent, at least in part, the complex interactions altered in AD, while being raised in uniform environments and with known genotype alterations. In particular, we aimed to understand the relationship between vulnerable brain circuits and memory deficits measured in the Morris water maze, and we tested several predictive modeling approaches. We used *in vivo* manganese enhanced MRI traditional voxel based analyses to reveal regional differences in volume (morphometry), signal intensity (activity), and magnetic susceptibility (iron deposition, demyelination). These regions included the hippocampus, olfactory areas, entorhinal cortex and cerebellum. The properties of these regions, extracted from each of the imaging markers, were used to predict spatial memory. We next used eigenanatomy, which reduces dimensionality to produce sets of regions that explain the variance in the data. For each imaging marker, eigenanatomy revealed networks underpinning a range of cognitive functions including memory, motor function, and associative learning, allowing the detection of associations between context, location, and responses. Finally, the integration of multivariate markers in a supervised sparse canonical correlation approach outperformed single predictor models and had significant correlates to spatial memory. Among a priori selected regions, expected to play a role in memory dysfunction, the fornix also provided good predictors, raising the possibility of investigating how disease propagation within brain networks leads to cognitive deterioration. Our cross-sectional results support that modeling approaches integrating multivariate imaging markers provide sensitive predictors of AD-like behaviors. Such strategies for mapping brain circuits responsible for behaviors may help in the future predict disease progression, or response to interventions.

1. Introduction

A key question in Alzheimer's disease (AD) research is how pathology differentially and sequentially affects vulnerable brain circuits, thereby giving rise to behavioral changes. Although critically important, detailed descriptions of interactions between genes and structural and functional phenotypes are poorly described. However, these interactions dictate the vulnerability for cognitive dysfunction in the context of aging and AD related pathologies. Investigating the circuits and mechanisms underlying cognitive dysfunction is important for understanding what triggers the switch from normal aging to AD, what predicts rates of disease progression, and how patient-specific therapeutic strategies may be developed [1]. Therapies for neurodegeneration have been directed to individual targets, such as altered synaptic transmission, amyloid deposition, or abnormal tau phosphorylation - all well-demonstrated pathologies in AD [2] [3]. Additional factors, such as cardio vascular status [4] [5], insulin resistance [6], and diabetes [7] may also influence cognition [4]. Importantly, neuro-immunological mechanisms, interacting with systemic inflammatory mediators and obesity, are thought to also modulate AD pathology [5, 8, 9]. Since attacking AD pathologies separately has not yet provided effective strategies for prevention or reduction of cognitive damage, we need models that provide an integrated view of how multiple variables and risk factors contribute to system wide dysfunction. We currently lack the quantitative integrative models required to understand this multifactorial condition.

To help understand the causative links between the biological and cognitive substrates typical of AD, it is helpful to conceptualize the brain as a set of interacting regions forming a spatially distributed network [10]. Structural networks integrate effects from changes occurring at different scales (synapse, cells, circuits), which in turn modulate the properties of functional networks. Several large-scale networks have been mapped in the brain and characterized by distinct functional profiles, such as sensory perception, movement, attention and cognition [11] [12]. However, it is not well understood how brain sub-networks map to the cognitive domain. Understanding these relationships in the normal brain, and their alterations in disease may inform on the mechanisms underlying mild cognitive impairment (MCI) or dementias such as AD [13].

One strategy to help understand how circuits influence behavior is to link imaging to the clinical AD cognitive phenotypes. The progressive loss of cognitive memory is commonly diagnosed using tests such as the Mini–Mental State Examination (MMSE) [14], the Montreal Cognitive Assessment [15], and others [16]. Clinical populations of individuals diagnosed with AD have shown overlaps in the patterns of gray matter atrophy [17], Aß distribution [18], and axonal density changes [19] [19], but there are also marked differences in brain atrophy [20] or tau pathology distribution [21]. Such differences may relate to population heterogeneity in terms of genetics, disease stage, or comorbidities. An alternative hypothesis to explore AD etiology is based on selective vulnerability of cells and axonal pathways favoring disease propagation. Imaging can provide *in vivo* biomarkers [22] [23] that are related to pathology as observed *ex vivo,* or to functional changes. To successfully link imaging to clinical phenotypes, we need to develop integrative models that explain the initiation, potentiation, and propagation of selective vulnerability in cells and networks that underlie AD processes, in relation to risk factors.

Neuroimaging approaches to map brain levels of behavioral descriptors have traditionally used voxel based statistical analyses (VBA) of deformation fields, structural and functional connectivity maps, vascular perfusion, or amyloid deposition and tau maps. But statistical approaches that pursue a dichotomous strategy, and aim to separate data according to image features do not necessarily explain the behavioral changes, nor do they disclose the biological processes underlying them. More recently predictive modeling approaches have been proposed to provide statistically relevant imaging correlates of memory changes spanning a continuum range, as observed in AD [24] [25] [26].

In this study, we have used mouse models for AD to develop such predictive approaches. Mice provide tools for dissecting the contributions of genes on circuits and behavior. In particular, they provide homogenous populations, and can be tightly controlled for genetic and environmental factors, thus simplifying the problem of mapping brain circuits responsible for behavior. To establish and test a novel integrative predictive modeling approach we have used the *APPSwDI$^{+/+}$/mNos2$^{-/-}$* (CVN-AD) strain. This mouse provides an AD-like pathological background required for studying the underlying mechanisms of regional vulnerability. CVN-AD mice replicate multiple AD pathologies, including amyloid and tau deposition, neuronal loss, altered microglial activity with typical AD-like inflammatory patterns and deficits in memory and learning [27] [28] [29] [30] [31]. The appearance of cognitive deficits with aging in this strain mimics processes in humans with AD and can be assessed using the Morris water maze test. This behavioral test is commonly used to quantify the loss of spatial learning and memory in animal models of aging and AD [32, 33].

To map behavioral changes to specific brain regions and networks we have used manganese enhanced *in vivo* MRI. Manganese ions ($Mn^{2+}$) are paramagnetic and induce T1 shortening [34], enhancing tissue contrast [35]. $Mn^{2+}$ has also been used to characterize trans-synaptic connectivity and axonal transport properties in rodents [36, 37] [38]. Importantly, $Mn^{2+}$ enters neural cells via voltage gated calcium channels and vesicular reuptake, presenting an alternative for task based fMRI in rodents [39] [40], while alleviating limitations due to the types of tasks that animals can perform in the magnet, or to anesthesia [41]. These strategies to characterize brain structure and function in CVN-AD mice in relation to age matched controls can identify vulnerable brain circuits responsible for behaviors typical of AD.

To identify vulnerable regions and networks that predict deficits in memory and learning, we used an integrative approach that was not linked to a single identified neuropathological mechanism, but was reflective of multiple concomitant factors. We evaluated how traditional mass-univariate analyses can predict behavior dysfunction, and followed with a multivariate approach involving dimensionality reduction. Eigenanatomy, a sparse dimensionality reduction method, was incorporated to extract brain regions responsible for changes in morphometry, signal intensity due to $Mn^{2+}$ uptake (reflective of brain activity), and magnetic susceptibility (reflective of altered iron homeostasis and conducive to oxidative stress and inflammation) [42] [43]. However, these methods analyzed individual biomarkers separately. We have employed both a data-driven as well as a hypothesis-driven approach to associate imaging phenotypes with behavioral markers for cognitive status and to identify circuits vulnerable to AD like pathology. Eigenanatomy produced candidate regions and circuits, and we selected regions that appeared important based on one or more biomarkers, and confirmed as relevant to AD from previous studies. To predict cognitive dysfunction based on *in vivo* multivariate imaging markers we used sparse canonical correlation analysis (SCCA) [44, 45]. SCCA selects regions so to maximize correlation among imaging and cognitive measures, in a supervised approach. The result is a network of regions that underlie changes in cognition, incorporated in a multivariate analysis. Our results provide insight into the relationships between structural networks and cognitive function in animal models of AD, supporting the value of multivariate approaches for humans with AD.

2. **Materials and Methods**

To test the hypothesis that we can identify vulnerable brain circuits involved in behaviors where aged mouse models for Alzheimer's disease (AD) differ relative to their age matched controls, we used *in vivo* multivariate magnetic resonance imaging (MRI) and the Morris Water Maze test for spatial memory. Our strategy examined each biomarker at a time, as well as an integrative predictive modeling framework.

*2.1 Animals*

The study was conducted under protocols approved by the Duke IACUC. CVN-AD (*APP*SwDI/*mNos2*[-/-]) (11 mice) and *mNos2*[-/-] controls (13 mice), aged 75.9 ± 4.4 weeks were handled and acclimated to water for 5 days. Mice were then implanted with Alzet 1007D minipumps (Durect Corp, Cupertino, CA), containing 100 µl of 64 µm/µl $MnCl_2*4(H_2O)$ (Sigma Aldrich, St Louis, MO), in 100 mmol bicine (Sigma-Aldrich, St Louis, MO). Animals were acclimated for 3 days following pump implantation, before behavioral testing. A summary of the experimental design is shown in **Fig. 1**.

*2.2 Behavioral Testing*

To test spatial memory which declines in AD, we used the hidden platform Morris Water Maze. For 5 days, mice received 4 trials a day (of maximum 60 s each), in trial pairs (60 s between trials) separated by 60 min intervals. Probe trials were run 60 minutes after the last trial on days 3 and 5. Swim time and distance were measured using Ethovision (Noldus Information Technology, Blacksburg, VA). Statistical analysis used SPSS (IBM, Armonk, NY), and included a repeated measure ANOVA, followed by Bonferroni corrected posthoc tests. A p-value < 0.05 was considered significant.

*2.3 Imaging*

*In vivo* imaging was done using a 7 T, 20-cm bore Bruker BioSpec 70/20 magnet (Bruker Biospin, Ettlingen GE/USR, Billerica, MA), interfaced to an Avance III console. The scanner has actively shielded gradients with integrated shims**.** The 198/114 mm outer/inner diameter insert gradient coil can supply 440 mT/m, at a rigif and rise time of 110 µs. We used a quadrature radio frequency transmit-receive cryogenic coil.

Two protocols were used to image the mouse brain at 100 µm resolution. To quantify morphometric changes and manganese uptake we used a T1-weighted RARE sequence with FOV 2x2x1 cm, matrix 200x200x100, NEX=2, BW=100 kHz, min TE=10.3 ms, TEeff=20.6 ms, TR=150 ms, echo spacing 10.3 ms, RARE partitions 4, acquired in 23 min. To estimate quantitative susceptibility maps we used a multi echo GRE with FOV 1.92x1.92x0.9; matrix 192x192x90, 8 echoes with spacing of 5.5 ms, NEX=1, TE1=3.9 ms; TR=100 ms, flip=30°, BW=62.5KHZ, respiratory gated, acquired in ~30-40 mins.

To ensure reproducible positioning, mice were restrained in a cradle equipped with ear bars, and a nose cone for isoflurane delivery (1.5±1%). Animals were monitored throughout the experiment, temperature and respiratory rate being maintained at physiological levels (37°C, 70-110 /min) by circulating warm water under the cradle, and adjusting the anesthesia level.

*2.4 Image Preprocessing*

Images were bias field corrected [46], skull stripped [47], and the resulting brain masks were manually edited. All brain images were rigidly aligned into the Waxholm space [48], and averaged to create a minimum deformation template [49] [50]. The template was labeled with a set of 332 regions, symmetric for the left and right hemispheres [51], defined on a single mouse brain [52]. To estimate morphometric differences, we used the log-transformed Jacobian determinants of the deformation fields (logJAC), mapping individual brains to the average. This enables comparisons with a symmetric distribution, with the same prior probability for shrinkage or expansion [53]. We used the T1-weighted images, where the voxel intensity reflects $Mn^{2+}$ uptake, which happens at least partly through calcium channels, to provide estimates of neuronal activity independently of hemodynamic activity [54]. Average T1-weighted image values for each mouse brain were normalized to a reference brain value. We calculated quantitative susceptibility maps (QSM) sensitive to iron, amyloid accumulation, and myelin using STI Suite [55]. Susceptibility values were directly used for comparison without referencing to any selected region of interest, which essentially sets the susceptibility reference to the mean susceptibility of the whole structure within the FOV. STI Suite uses a Laplacian-based method to unwrap the phase, after which the background is eliminated using vSHARP (Wu et al., 2012). The corrected phase images were combined, weighting the two channels. We used a two-step streaking artifact reduction regularized reconstruction (STAR-QSM) [56], which optimally weighs data consistency and smoothness for both high and low susceptibility variations.

*2.5 Voxel Based Analysis*

All three contrast images (log Jacobian, normalized T1-weighted RARE, and QSM) were mapped into the space of the minimum deformation template, and smoothed with a 200 µm kernel. This kernel was selected to highlight the scale of features at which we expect to detect pathology-related differences across these MRI-based measurements. SPM [57] was used with voxel-wise false discovery rate correction.

*2.6 Eigenanatomy*

We hypothesized that eigenanatomy [25], a sparse dimensionality reduction technique, will confer increased power to detect differences between genotypes. The method approximates an eigen decomposition of an image set with spatial basis functions (eigenanatomy vectors) that are unsigned, sparse, and anatomically clustered. We employed the eigenanatomy vectors as anatomical and functional imaging predictors. We leveraged the technique in two stages: first, to reduce the dimensionality in an unsupervised setting; second, to perform a supervised regression within the setting of cross-validation.

The goal of eigenanatomy is to identify sparse functions which approximate the eigenvectors ($v^p = X^T v^n$ and $v^n = X v^p$) of the $n \times p$ image matrix $X$, where each row $n$ represents the data for one subject, and has $p$ entries (voxel values). The method approximates $X$ with $i$ sparse singular vectors $v^{sp}_i$, treating the positive $\left(v^{sp}_i{}^+\right)$ and negative components ($v^{sp}_i{}^-$) separately, and imposes positivity constraints on both:

$$\underset{v^{sp}_i{}^+, v^{sp}_i{}^-}{argmin} \left\|X^T X v^{sp}_i{}^+ - v^p_i{}^+\right\|^2 + \left\|X^T X v^{sp}_i{}^- - v^p_i{}^-\right\|^2; \text{subj. to } \left\|v^{sp}_i{}^+\right\|_1 = \left\|v^{sp}_i{}^-\right\|_1 = \gamma \text{ (Eq.1)}$$

where $\gamma$ is the sparseness parameter. The weight on the L1 penalty is set to reach the desired sparseness and guarantees that only a subset of voxels is considered [44] [58]. The minimization uses a nonlinear conjugate gradient, with sparseness imposed through soft thresholding S(v, $\gamma$), which rejects clusters below a threshold. The resulting pseudo-eigenvectors are sparse, unsigned, and represent the input data (X) through weighted averages.

To estimate the generalization performance for our models we divided the data in training and testing groups (75%, and 25%), and we used a 4-fold cross-validation, as described in [59]. We assigned data to training (n=18), and validation (n=6) partitions in a balanced way using caret (caret.r-forge.r-project.org/), then evaluated the average root mean square error (RMSE) over the test partitions. Dimensionality reduction was performed using the function *sparseDecom* in ANTsR (https://github.com/ANTsX/ANTsR) for each of the imaging contrasts (logJac, T1w, QSM), producing areas that have maximum covariance between subjects (Cook et al. 2014). For each imaging contrast we used 2 eigenvectors, a sparsity threshold of 5%, and a min cluster size of 250 voxels, or 6 eigenvectors for all three imaging contrasts used in the multivariate approach. These eigenregions were projected against the training set to generate statistical models that predict swim distance at day 4. The summary of the statistical models performance applied to the testing sets included the root mean square error (RMSE), Pearson correlation, adjusted $R^2$ and p value (considered significant at $p<0.05$).

*2.7 Prior-Based Prediction Using Sparse Canonical Correlation*

We used an anatomically informed, prior-based approach to constrain the solution space, to test whether specific brain regions were associated with cognitive performance. These regions were determined based on: 1) eiganatomy results, producing a supervised decomposition, and 2) anatomical priors on regions known for their involvement in AD, and which are recognizable in mouse models. These regions included the hippocampus, parasubiculum, and fornix [60] [61] (Micotti et al. 2015) (Badea et al. 2016), as well as the entorhinal and motor cortices. Each of these regions is likely to incur changes during training in the water maze. These regions were used to initialize SCCAN [58] to infer their influence on behavior.

We hypothesized that we could identify brain networks associated with behaviors based on canonical correlation (CCA) [62], the multivariate extension of correlation analysis. To find the linear projections of two random vectors, CCA maximizes the correlation between the two linear combinations of the variables in each data set. Our datasets consist of imaging (*X)* and behavioral parameters (*Y)* for *n* subjects. The imaging parameter is a large $nxp$ multidimensional matrix, and the behavioral parameter is $nxq$; where *p* is the number of voxels in *X* for each subject, and *q* is the number of behavioral parameters in *Y* for each subject.

Due to the greater size of the imaging matrix compared to behavior, CCA becomes ineffective. Instead sparse CCA has been used as a dimensionality reduction method to produce the solution vectors, *x (p ×1)* and *y (q × 1)*, which act as weights on columns of *X* and *Y* [58].

$$x^*, y^* = \underset{x,y}{argmax} \frac{xX^TYy}{\|Xx\| \|Yy\|} ; subject\ to\ \sum_j \|x^j\|_1 \leq s,\ x^j \geq 0\ (Eq.2)$$

The *x\** solution vector is subjected to the "L$_1$" norm, $\|.\|_1$, producing non-zero entries below the chosen sparsity threshold (Kandel et al. 2013). The gradient of the objective function in (Eq.2) is calculated with respect to x and y. A nonlinear gradient descent optimizer is used to provide the solution vectors. Additionally, a cluster threshold and smoothness constraint are enforced to retain anatomically meaningful brain regions. As a result, sparse CCA produces solution vectors *x\** and *y\**, dimensioned as subsets of variables that maximize the correlation between imaging and behavior.

Here we focused on a single vector to represent behavioral performance, i.e. swim distance at day 4 (SD4). We used SCCAN to perform a sparse regression between imaging and behavior.

$$\underset{x}{argmin} \frac{1}{2} \|Xx - y\|_2^2 + \frac{\lambda_1}{2}\|x\|_2^2 + \frac{\lambda_2}{2}\|\nabla x\|_2^2; \ subject\ to\ \|x\|_1 \leq s \ (Eq.3)$$

where *s* is the desired sparseness, $\lambda_1$ a ridge penalty which alleviates the problem of multicollinearity amongst regression predictor variables, and $\lambda_2$ a smoothness penalty. This is solved through a projected gradient descent (Kandel et al. 2013). We note that in recent approaches the L0 penalty, which guarantees that only a subset of voxels is considered for the model, has been replaced with the convex approximation given by an L1 penalty, which yields virtually identical results, in a robust approach to regression [63](Kandel et al. 2013). The soft thresholding operator was used to update the sparse projection of the solution vectors at each step of the optimization.

We used SCCAN for a positively constrained optimization that finds projection vectors in the minimum deformation template space, which maximizes the relationship between image markers (separate contrasts, and in combination) and swim distance at day 4. The inputs for SCCAN included the training set for both imaging and swim distance (SD4) through the function *sparseDecom2* (Dhillon et al. 2014).

Both the imaging and behavioral data were split into training and testing sets in a ratio of 75:25, which allowed for 18 variables/subjects for training and 6 for testing. We chose the swim distance on day 4 (SD4) as the behavior to be predicted based on its robust ability to separate the groups. The sparse canonical correlation was initialized from each of the *a priori* selected image regions (1 eigenvector), using a medium prior (0.5). We used sparseDecom2 with 5% sparseness, 250 voxels cluster threshold, 15 iterations. No sparsity constraints were enforced on behavior. The projection vectors obtained from the training set model the relationship with behavior. These models were then used to predict swim distance for the testing set. To assess the quality of the predictive modelling approach we determined the relationship between the predicted and measured SD4.

We examined how the solution vectors from sparse CCA of different imaging contrasts performed as predictors on unseen behavioral data. We aimed to identify brain regions with a reliable association between each imaging marker and behavior using sparseDecom2, with a sparseness value that selects a small, informative subset of voxels (250 voxels, and 5% sparsity). Finally, the resulting eigenregions were used to fit a linear model to SD4.

We tested several models to predict behavior based on imaging biomarkers using a 4-fold cross validation scheme. First, we examined how the clusters surviving FDR correction in the voxel based analyses (VBA) in the three different imaging contrasts performed as predictors for behavioral data. We compared the performance of these models with those generated from the sparse decomposition (SD) for each of the imaging contrasts. Finally, we fused the information from all imaging contrasts in a behavior supervised sparse decomposition that maximized the canonical correlation between imaging and swim distance. This approach was used in a hypothesis generating mode throughout the whole brain to identify brain circuits responsible for AD like cognitive decline, as well as to test our hypotheses for regions expected to be involved in memory or motor function.

All prediction methods were executed using R (www.r-project.org) and the ANTsR package (http://stnava.github.io/ANTsR/). The compute times required for each fold in our optimization experiments ranged from 2 minutes (for 2 vectors, 5% sparsity) up to 2 hours and 22 minutes (for 50 eigenvectors, 5% sparsity), using a MacPro equipped with 12 core 2.7 GHz Xeon E3 processors, running 10.13.6 Mac OS Sierra.

FSLEyes (https://fsl.fmrib.ox.ac.uk/fsl/fslwiki/FSLeyes) was used for visualization of statistical parametric maps overlaid on the average T1W MEMRI template generated from $mNos2^{-/-}$ controls.

## 3. Results

To help uncover the relationship between cognition and the biological substrates underlying Alzheimer's disease (AD), experiments were carried out using a well-characterized mouse model of AD, to reduce genetic and environmental diversity. Behavior and imaging data were subjected to univariate and multivariate analyses.

Spatial memory was examined through acquisition performance in the Morris water maze **(Fig. 2)**. Swim distance to the hidden platform declined across testing for both genotypes (**Fig. 2A**). Significant genotype effects emerged on block trials 3-5, where swim distances decreased in the $mNos2^{-/-}$ control mice relative to the CVN-AD strain (p≤0.05). Within genotype, swim distances decreased from trial 1 to trials 3-5 in $mNos2^{-/-}$ control mice (p≤0.002), whereas in CVN-AD animals this parameter only declined from trial 1 to 5 (p=0.005). A similar relationship was observed for swim time (**Fig. 2B**). Here, swim times were reduced significantly on trials 3 and 4 for the $mNos2^{-/-}$ compared to CVN-AD animals (p≤0.025). Within genotype, swim times in $mNos2^{-/-}$ animals declined from trial 1 to trials 3-5 (p<0.001). By comparison, in CVN-AD mice swim time was decreased only from trial 1 to 5 (p=0.001).

Analysis of learning performance on the probe trials for swim distance on days 3 and 5 of acquisition testing revealed that $mNos2^{-/-}$ control mice had already identified the northeast (NE) quadrant as the target zone, whereas CVN-AD mice had not made this distinction (**Fig. 2C**). Swim distances in CVN-AD animals on day 3 were shorter in the NE target (p=0.026) and longer in the northwest (NW) and southwest quadrants than $mNos2^{-/-}$ controls (p≤0.018) (**Fig. 2C,** *left*). On probe day 5, $mNos2^{-/-}$ animals maintained increased swim distances in the target quadrant, whereas in CVN-AD mice swim distances were longer in the NE and northwest than the other quadrants (**Fig. 2C**, *right*). On this probe trial, the only genotype difference was in the northwest quadrant where CVN-AD mice swam over a longer distance than $mNos2^{-/-}$ controls (p=0.003). Similar relationships were observed with swim time (**Fig. 2D**). With the probe trial on day 3, $mNos2^{-/-}$ mice swam for longer times in the NE target quadrant (p=0.002) and shorter times in the northwest (p=0.004) with a trend for the southwest (p=0.064) quadrant compared to the CVN-AD animals (**Fig. 2D**, *left*). By day 5 probe trial, swim time remained augmented in the northwest quadrant for the CVN-AD mice (p=0.045) (**Fig. 2D**, *right*). These results indicate that the CVN-AD mice had learned that the north quadrants contained the target, but they were unable to discriminate the NE from the NW quadrant. Collectively, these results demonstrate that acquisition duration in the Morris water maze is prolonged in CVN-AD mice, which fail to identify the NE quadrant as the target zone in the probe trials. Since in the acquisition test genotype differences were most robust on trial 4 for swim distance, and provided a clear separation of the two genotypes, these data were used as the dependent variables for predictive modeling.

To characterize CVN-AD mice relative to age matched controls we used manganese enhanced imaging (MEMRI). We developed two imaging protocols to characterize anatomy and memory function based on: 1) the log-jacobian of the deformation fields; 2) the T1-weighted signal intensity (normalized to the average value determined for a reference brain) reflecting manganese accumulation; and 3) quantitative susceptibility maps (QSM). Representative images for one animal are shown in **Fig. 3A**. We constructed minimum deformation templates (MDT) as study specific population atlases for each of these contrasts (**Fig 3. B**), and derived Jacobian and QSM maps (**Fig. 3C**). All subsequent analyses were performed in the common space of the MDT.

We sought brain networks for which changes in imaging markers explain changes in behavior. Using regional and voxel based analyses (VBA) we identified significant changes in volume (Jacobian of deformation fields), manganese accumulation (normalized T1-weighted signal intensity), and quantitative susceptibility maps (QSM). All image contrasts identified differences in the olfactory areas, hippocampus, entorhinal cortex, and cerebellum, and the volume and T1w analyses revealed a role for the retrosplenial cortex (**Fig. 4**).

The reductions in volume in olfactory areas, thalamus, and hippocampus were accompanied by increased magnetic susceptibility, and lower manganese uptake (**Fig. 4 A-D**). Areas of the hypothalamus, entorhinal cortex, hippocampus and subiculum showed decreased susceptibility. Increases in susceptibility were noted in the caudate putamen and red nucleus. T1w signal intensity was overall lower on CVN-AD mice, and local differences between genotypes were significant at FDR 0.2 in the olfactory areas, motor cortex, primary somatosensory (S1) cortex, hypothalamus, and hippocampus, in particular the dentate gyrus and subiculum, and cerebellum. A more extensive presentation of the regional results can be found in the **Supplementary Tables**.

Separate analyses for volume, T1w signal and QSM were conducted using the clusters with significant VBA differences between the two groups to model the relationship with behavior (**Fig. 4 E**). Volume was ranked as the best predictor, followed by T1w signal, then QSM. The root mean square error (RMSE) ranged from 259±57 cm for clusters with significant atrophy, to 274±54 cm for clusters with decreased T1w signal, while the largest RMSE was obtained for the combined QSM clusters (290 ±51cm). **Table 1** reports the performance of all tested methods to produce the final models based on the 4-fold cross validation (testing) RMSE, the correlation and variance explained by each method, and their rank in terms of predictive performance.

We then used eigenanatomy to identify areas of the brain with maximum covariance between subjects (Cook et al. 2014), producing a sparse decomposition (SD) for each of the imaging contrasts. The eigen regions based on volume revealed a network comprised of the frontal pole, septum, medial thalamus, retrosplenial and cingulate cortex, amygdala, the CA1 area of hippocampus, and fimbria. In addition, we noted the involvement of sensory and motor cortices, and of the caudate putamen (**Fig. 5A, top row**).

The T1w signal intensity-based decomposition revealed a network including olfactory areas, the cingulate cortex, bed nucleus of stria terminalis, substantia innominata, amygdala and hippocampus (**Fig. 5B, top row**).

The QSM based decomposition revealed involvement of the olfactory areas, septum, and cingulate cortex, the hippocampus, bed nucleus of stria terminalis, substantia innominata. Iron rich regions were also involved such as the substantia nigra, caudate putamen and globus pallidus (**Fig. 5C, top row**).

By adding elements of supervision through the sparse canonical correlation of imaging and behavior we observed additional areas relative to the networks identified before. These included the fornix, dorsal thalamic nuclei, as well as the ventricles for the volume based decomposition (**Fig. 5A, bottom row**); the medial thalamic nuclei for the T1W based decomposition (**Fig. 5B, bottom row**); white matter such as the anterior commissure and corpus callosum (including areas below the motor cortex and S1) for the QSM based decomposition, and more extensive areas on the ventral hippocampus and primary somatosensory cortex (**Fig. 5C, bottom row**).

Specific brain areas (eigen regions) found to be common in the results for all three contrasts included the olfactory, cingulate cortex and retrosplenial cortex, hippocampus, as well as the motor cortices, and septum. The substantia innominata and bed nucleus of stria terminalis were common to T1w, and QSM.

Using a supervised approach and the combined regions from volume, T1w and QSM provided the lowest RMSE of 233.08 ±101.78 cm, a significant correlation of 0.94, explaining 88% of the variance in the swim distance (**Fig. 5D**). The tightest confidence intervals and largest adjusted R2 were obtained using the combined imaging biomarkers in the supervised approach (**Fig. 5E**, and **Table 1**).

A nonparametric Kruskal Wallis test did not reveal a significant difference for the generalization performance estimated based on the 4 fold cross-validation (testing RMSE) for all models (p=0.09, chi square=22.9, df=15). To evaluate the 16 models performance, we compared their predictions with the measured values for the swim distances for the full data set. The Kruskall–Wallis analyses were followed by posthoc Tukey Kramer tests to control for the family wise error rate. Our results (**Table 2, Supplementary Figure 1**) indicated that the supervised multivariate approach outperformed 12 of the other models (p=6.9*10$^{-5}$, chi=45.3, df=15). These analyses also indicated that amongst the imaging contrasts studied, volume was the best predictor for behavioral performance in the water maze, followed by QSM, and then T1w.

To test whether structures about which we have previous hypotheses are predictive of behavioral performance in the Morris water maze, we selected regions that appeared significant in one or more of the single biomarker analyses. These regions play a role in spatial memory and included the hippocampal formation, entorhinal cortex, parasubiculum, fornix, the primary and secondary motor cortex. **Fig. 6** shows the predictive correlation for these models. The root mean square error, and significance of the models predicting relationships between imaging and behavior generally improved or were similar to the best predictor when the model included all three factors (**Table 3**). However, the single region analysis underperformed relative to the whole brain supervised sparse decomposition (SSD or SD2) analysis, with a maximum correlation of 0.76 for the parasubiculum in the combined analysis (explaining 56% of the variance), and 0.73 for its volume (explaining 51% of the variance). This suggests that our animals model a complex, network- rather than a region-based disease.

Our results indicate that multivariate, integrative predictive modeling approaches may outperform any single one imaging modality, giving us the ability to map vulnerable brain circuits responsible for cognitive changes. This is in particular important for a multifactorial disease like AD, where the same regions are affected by multiple pathologies.

### 4. Discussion

Neurodegenerative conditions such as AD arise from multifactorial pathological processes. Integrative modeling is an important step towards better understanding this complex disease etiology, as well as predicting its trajectory. Recent efforts to produce models for disease progression and response to treatment have shown promise in AD patients, and cognitively normal people at risk for AD [64]. However, the genetic variability and differences in environmental conditions to which patients have been exposed make these studies difficult. Animal models provide attractive tools for conceptually advancing our understanding of complex pathological processes, and factor to factor interactions. Moreover, animal models are required for testing therapeutic interventions. Hence we used a mouse model [27], previously characterized using pathology, behavior [28], and *ex vivo* diffusion tensor MRI [65], and whose development of cognitive dysfunction mimics the development of AD.

Our results support previous findings on learning and memory deficits in mouse models of AD [66], [67, 68]. Such deficits have been associated with the presence of amyloid [68, 69], tau [70], altered synaptic plasticity [71], or to inflammation and neurodegeneration [66] [28] [30]. The impairment in the acquisition of the Morris water maze was evident for CVN-AD mice relative to *mNos2$^{-/-}$* controls when both swim distance and swim time were analyzed. These deficiencies were verified in the probe trials where CVN-AD mice failed to discriminate the target quadrant from the other quadrants early in testing and from the northwest quadrant at the end of testing. Our results thus confirmed that CVN-AD mice were impaired in spatial memory, in agreement to previous publications [28] [31]. Moreover, we identified swim distance during the 4$^{th}$ day of trial test as a robust measure that allows clear differentiation of CVN-AD models from age matched *mNos2$^{-/-}$* controls.

To bridge between structural and functional imaging correlates, we used manganese, which modifies MRI signals and accumulates during behavioral training and testing. Manganese increases image contrast due to differential uptake by various brain areas [72], [73, 74]. We have used this property to more accurately estimate morphometric changes *in vivo.* T1 shortening occurs as a consequence to intracellular manganese uptake, upon neuronal excitation/depolarization. These processes are known to correlate with synaptic firing [75] [76]. Thus, manganese enhanced MRI (MEMRI) provided a functional mapping tool [38] [37] [40] [77], and has been proposed as a surrogate to fMRI, which relies on hemodynamic changes. Our use of MEMRI is equivalent to a spatial memory -based fMRI, with the limitation that it lacks the temporal resolution. Rather MEMRI presents a picture of the brain activity, after integration over the duration of the behavioral tests. Importantly, MEMRI bypasses the shortcomings of anesthesia on task performance and we showed that MEMRI can be used to map complex brain circuits involved in spatial memory.

The observed decrease in MEMRI signal after behavioral testing supports the loss of neuronal activity in memory circuits in old CVN-AD mice. A loss of MEMRI signal has also been demonstrated in mouse models of tauopathy [40, 78]. In contrast, MEMRI signal was shown to increase 24 h after MnCl2 injection in young 5xFAD mice [79]. This model of AD demonstrates abundant amyloid deposition generated from a high level of overexpression of mutated human amyloid precursor protein [80]. In addition to the differences between mouse models, the differences between these studies results may be explained by two effects: 1) high neuronal activity was present in younger mouse models of AD, and later decreased with age, and/or 2) our study measured a functional effect related to maze training, specific to memory function and diminished in older models of AD.

The T1w MEMRI images suggested decreased Mn movement into the cells in CVN-AD models, in particular in olfactory areas, the hippocampus, hypothalamus, motor cortex, and also the cerebellum. The lower manganese uptake in CVN-AD mice was accompanied by atrophy in the olfactory areas, thalamus, and hippocampus, and was also associated with increased magnetic susceptibility. However, a reduced brain activity was expected to cause susceptibility reduction. Indeed, several areas of the hippocampus, entorhinal cortex and parasubiculum, as well as hypothalamus had lower susceptibility values in AD models. A decrease in susceptibility can be attributed to reduced manganese uptake, and also to diamagnetic properties of amyloid, or myelin. QSM changes did however overlap with areas of volume reduction, possibly associated with cellular and dendritic density reduction, leading to loss of anisotropy. QSM increases were noted in the hippocampus, olfactory areas, and red nucleus. VBA also identified QSM increased in iron rich areas, such as the globus pallidus and caudate putamen. Alterations in iron metabolism or the presence of microbleeds may lead to the observed predominant susceptibility increases. The increased susceptibility in areas such as the caudate putamen (Kirsch et al., 2009) and hippocampus was associated with decreased T1w values, thus could result from the higher level of iron from neurofibrillary tangles, and the aggregation of iron containing amyloid plaques. Hippocampal QSM increase has been associated with amyloid pathology in humans, moreover it was predictive of faster cognitive deterioration [81].

VBA revealed that morphometric changes survived the highest stringency in thresholded statistical maps, relative to QSM and T1-w signal reflecting manganese uptake. Significant changes in all three parameters were present in olfactory areas, septum, hippocampus in specific layers and the dentate gyrus, subiculum, hypothalamus and cerebellum. In summary, while the exact pathology underlying the observed changes remains unclear, these may be attributed to atrophy associated with neurodegeneration (or conversely, increases due to astrogliosis), to reduced manganese uptake and myelin loss; or to amyloid and abnormal iron deposition.

Among phenotypic differences assessed with VBA, volume ranked best for its ability to predict memory. We used eigenanatomy [25] to exploit the covariance in a dataset, and identify a reduced set of voxels, assembled into coherent regions. We then used regression in this reduced dimensionality space to identify associations with spatial memory. The regularized sparse decomposition identified that the volume of the cingulate and motor cortex areas had good predictive value for spatial memory. Additionally, the decomposition solutions based on T1w signal revealed that lower activity in the limbic thalamus, in the motor, cingulate and entorhinal cortices, as well as in the globus pallidus, were all predictive of the behavioral performance. Some of these areas were also present in the decomposition solutions based on susceptibility, which emphasized the role of the olfactory, motor, cingulate, and hippocampal areas.

While distinct regions were associated with behavior for different imaging parameters, a number of these regions were common between the three contrasts, and were overlapping with areas known to be involved in AD. These involved a network including septum, hippocampus, cingulate, retrosplenial and motor cortices. Our results also suggest a role for limbic thalamic nuclei, substantia innominata and amygdala. While the eigenregions were more extensive than the clusters obtained from voxel based analyses, these were also likely to be involved in multiple behaviors, such in motor planning and execution of the task, besides spatial memory.

Motivated by such hypotheses we have integrated multivariate imaging biomarkers into a modeling framework to map brain circuits that predict performance in the water maze. We selected regions identified as important in our previous analyses for one or more of the biomarkers, and with demonstrated involvement in AD. We found that gray matter regions such as hippocampal [82], parasubiculum [83], entorhinal[84], retrosplenial [85] and motor cortices provided good predictions for spatial memory performance. While our techniques were limited in resolution to 100 μm, we were able to identify the fornix volume, and QSM as predictors for spatial memory function. Other white matter tracts such as the anterior commissure (connecting the temporal lobe structures), and corpus callosum (largely responsible for interhemispheric connectivity) may need to be investigated in future studies. Such studies may help understand how disease propagation within brain networks leads to cognitive deterioration.

Finally, we assessed the value of individual and combined imaging biomarkers in predicting spatial memory for whole brain based analyses, and for selected regions which appeared as eigen solutions for one or more imaging markers and had demonstrated roles in AD. We found that the combined supervised approach improved the prediction accuracy relative to single biomarkers, in agreement with [25] [26] [86].

A limitation of our approach comes from the small sample size, a problem that is common to many preclinical studies. This restricts us to using simpler models. However, this is alleviated by the genetic similarity of mouse populations relative to clinical populations. While in this study we demonstrated a promising approach for predictive modeling in mouse models of AD, future studies would benefit from larger samples. Such studies may test other correlatives, and attempt to build genotype specific models.

Possible neurotoxic effects may limit manganese studies, in particular their applicability to longitudinal studies in animal models, and they are certainly not amenable to human studies. To maximize contrast while reducing toxicity, we chose a small but continuous delivery method, via implanted mini-pumps, over a single injection [87]. An additional consideration when using QSM in manganese dosed models of AD is the possibility of compound effects, e.g. iron accumulation, which is likely to dominate in effect size, and may induce oxidative tissue damage [88] [89]. However, our technique produced quantitative susceptibility maps with exquisite contrast by exploiting the MR phase, in addition to providing morphometry and T1-weighted signal information. This approach holds promise as a more direct measure of functional information based on imaging [90] [91].

Since imaging biomarkers may indicate changes before overt cognitive decline [92], such approaches can help with early diagnosis and patient stratification. It would be interesting to include vascular biomarkers, which may constitute early events in fronto-temporal dementia [92] and AD [93]. Future studies may address predictions along a temporal scale, and give insight into the dynamics of interactions among pathological factors, in relation to disease propagation.

To the best of our knowledge this is the first application of sparse predictive modeling integrating multivariate biomarkers for structure and brain activity, to map circuits responsible for behavioral dysfunction in models of AD. Mapping cognition to brain circuits will increasingly rely on such multivariate statistical algorithms involving clustering, module detection, or other dimensionality reduction approaches, which offer increased power to identify signatures of neurodegenerative disease. The success of such translational approaches will allow testing of mechanistic hypotheses using mouse models, and help develop better models for complex diseases.

In conclusion, we show that MEMRI can provide an integrated approach for studying brain dysfunction in rodent models of neurological disorders. Our approach synergized information from multivariate imaging and behavioral markers, allowing for observation of multifactorial biological processes and enabling future modeling of such factor-factor interactions, locally or as they spread over physical brain networks, to alter functional networks and ultimately cognition. Our results demonstrated that, compared to using information provided by individual biomarkers in isolation, an integrative approach can better predict cognitive outcomes, including memory deficits. Such multivariate approaches hold promise to help discover mechanistic links between the structural and functional components of brain circuits that underlie cognitive dysfunction in AD.

## Acknowledgments

Imaging was performed at the Center for In Vivo Microscopy, supported through P41 EB015897 (G Allan Johnson). We thank all CIVM-ers for their efforts to build and maintain this resource, and a collaborative learning environment. We thank Michael Lutz, Sayan Mukherjee and Nian Wang for helpful discussions, John Nouls for maintaining the 7 T magnet; Gary Cofer for generously sharing his MR knowledge with students, James Cook and Lucy Upchurch for supporting the computing resources. We thank Ramona Rodriguiz and Christopher Means for help with behavioral assays. The behavior testing was done on instruments acquired with support from the NC Biotechnology Fund. This work was supported by the National Institutes of Health through K01 AG041211 (Badea), R01 AG045422 (Colton), R56 AG051765 (Colton), R56 AG 057895 (Colton, Badea, et al.).

# Tables and Figure Legends

**Table 1.** Model performance comparison based on a whole brain unbiased analysis, showing the testing root mean square error of the predictions (RMSE); and the whole set based Pearson correlation (corr), associated p value, and the explained variance (adjusted R2).

**Table 2.** To evaluate the 16 models performance, we have compared predictions with the true values for the measured behavior parameters (measured in cm) based on the full data set. Our Kruskall–Wallis analyses were followed by posthoc Tukey Kramer tests to control for the family wise error rate. Our results indicate that the supervised multivariate approach outperforms 12 of the other models. CI: confidence interval. VBA: voxel based analysis; SD2: supervised sparse decomposition.

**Table 3.** Model performance comparison based on a ROI prior initialized based analysis, showing the testing root mean square error of the predictions (RMSE); and the whole set based Pearson correlation (corr), associated p value, and the explained variance (adjusted R2). Hc: hippocampus; PaS : parasubiculum; CEnt: caudal entorhinal cortex; fx: fornix; M1: primary motor cortex; M2: secondary motor cortex.

**Figure 1. Experimental timeline.** Each animal was implanted with a $MnCl_2$ filled minipump on day 0 (d0), then acclimated for 3 days. Behavior was assessed over 5 days (d3-d7) using the Morris Water Maze (MWM). Subsequently mice were imaged using *in vivo* MRI.

**Figure 2. Acquisition performance and probe trial results (mean± SEM).** (A) RMANOVA for swim distance detected a significant within subjects effect of time [$F(4,88)=22.436$, $p<0.001$] and a significant time by genotype interaction [$F(4.88)=2.451$, $p=0.012$]; the between subjects effect of genotype was significant also [$F(1,22)=13.229$, $p<0.001$]. (B) The RMANOVA for swim time noted that the within subjects effects of time [$F(4.88)=26.706$, $p<0.001$], the time by genotype interaction [$F(4,88)=2.667$, $p=0.037$], and the between subjects effect of genotype [$F(1,22)=5.466$, $p=0.029$] were significant. N=13 *mNos2$^{-/-}$* mice and N=11 CVN-AD mice. (C)

A RMANOVA for swim distance detected that the within subjects main effect of day was not significant, but the effect of zone [F(3,63)=9.693, p<0.001) and the zone by genotype interaction [F3,63)=4.618, p=0.006) were significant. There was a trend for the day by zone by genotype to be significant [F(3,63)=2.241, p=0.092]. No other within subjects interaction or the between subjects effect of genotype were significant. (D) A similar effect was found with swim time where the RMANOVA only detected the zone effect [F(3,63)=11.741, p<0.001] and the zone by genotype interaction [F3,63)=5.790, p=0.001] to be significant. NE=northeast, NW=northwest, SE=southeast, SW=southwest. One *mNos2<sup>-/-</sup>* control was not tested on the probe trial day 5 due to a lesion. N=12 *mNos2<sup>-/-</sup>* mice and N=11 CVN-AD mice; *p<0.05, *mNos2<sup>-/-</sup>* versus CVN-AD mice; +p<0.05, within genotype versus day 1 acquisition trials.

**Figure 3. T1-weighted RARE and mGRE images were acquired for 24 mice, and the mGRE was used to calculate quantitative susceptibility maps.** Representative images from one mouse are shown in (A). The control group was used to calculate minimum deformation templates (MDT) (B) for T1w RARE and mGRE images, to produce the deformation maps (C), and MDT for quantitative susceptibility maps (D).

**Figure 4. (A) Voxel-based analysis (VBA) identified genotype differences in all three biomarkers (volume, T1-weigted signal, and susceptibility).** Statistical t maps were thresholded using false discovery rate (q). Regional properties identified areas of significant differences for (B) volume; (C) T1w signal intensity normalized to total brain; (D) magnetic susceptibility (QSM). Local atrophy was evident in the olfactory areas, septum, and dentate gyrus of the hippocampus in the CVN-AD model. Manganese uptake was lower in these mice, with the exception for the septum/fornix. Susceptibility was increased in the olfactory areas, caudate putamen, and dentate gyrus of hippocampus, but decreased in the CA1, hypothalamus, entorhinal cortex, and subiculum. (E) VBA identified regions were used for predicting behavior, based on regions with either positive or negative effects for volume, T1W signal, and QSM.

**Figure 5 (A). Using volume (A), T1w (B), QSM (C), or combination of the three biomarkers (D) as predictors we identified significant associations between eigen regions and swim distance on day 4.** For all contrasts, the 1st row represents the networks identified by the sparse decomposition. The 2nd row represents the networks identified based on a behavior supervised decomposition. Panel E shows how the correlations between the predicted behavior trait for each of the models applied to the full data set, and the associated confidence intervals. Our results suggest the integrative approach has increased value when applied to a model of a complex, multifactorial disease such as AD.

**Figure 6: Relative to single biomarkers, the combined set of three contrasts performed in general better or just as well as the best predictor, in terms of the ability to predict swim distance for regions selected because they were *a priori* expected to be involved in memory or motor function, and or AD pathology.** We note the good predictions for the parasubiculum, involved in spatial navigation.

**Supplementary Material**

**Supplementary Table 1.** Volume Differences

**Supplementary Table 2**. T1w signal intensity differences

**Supplementary Table 3**. QSM differences in CVN-AD mice relative to *mNos2$^{-/-}$* controls

**Supplementary Figure 1**. Model performance comparison based on a whole brain unbiased analysis, showing the root mean square error (RMSE) between the predictions and the true values for the swim distance produced for the 16 models.

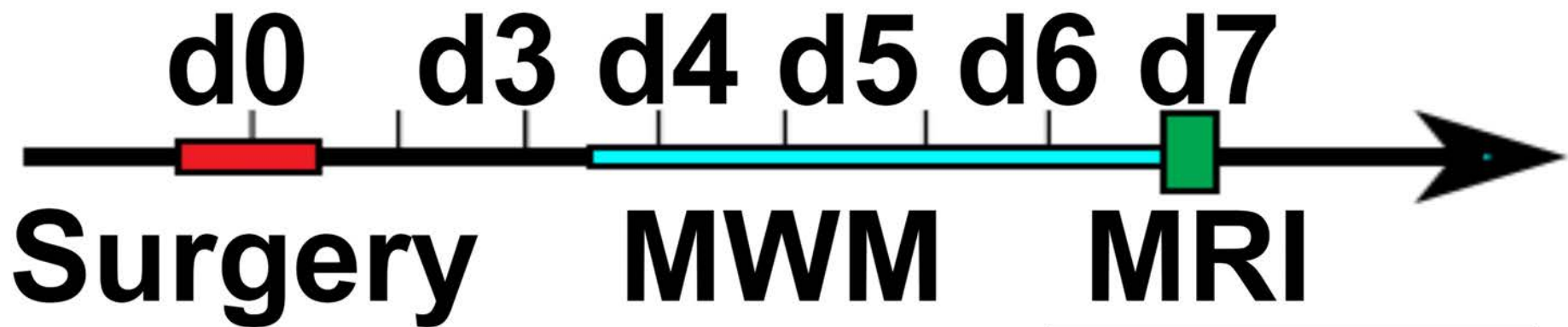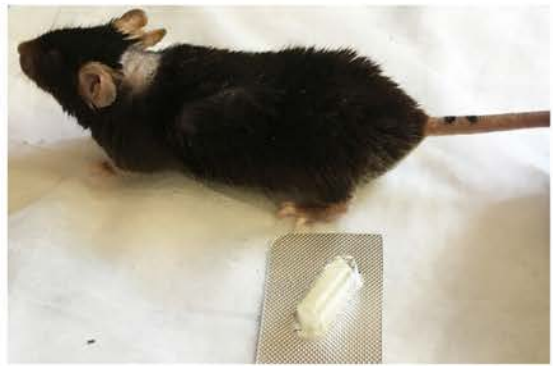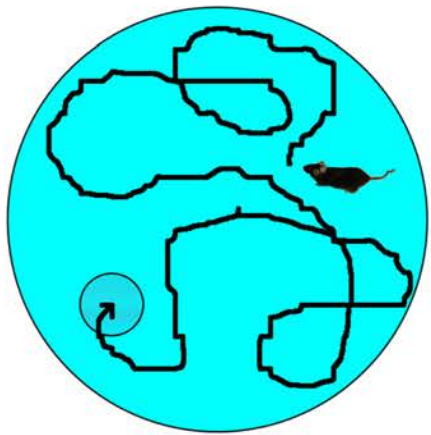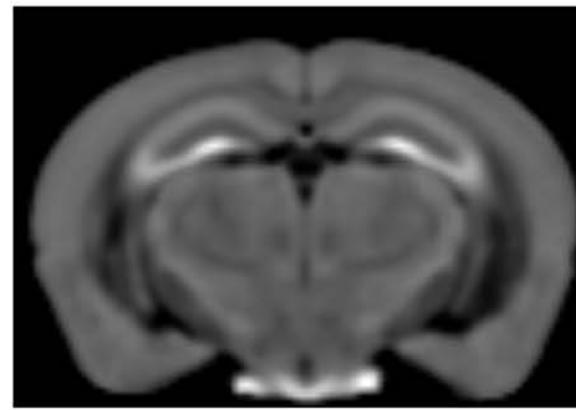

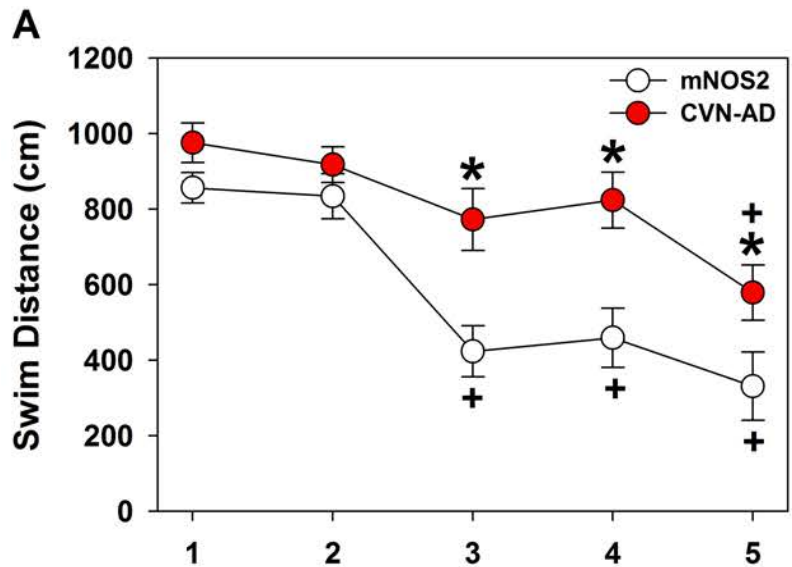
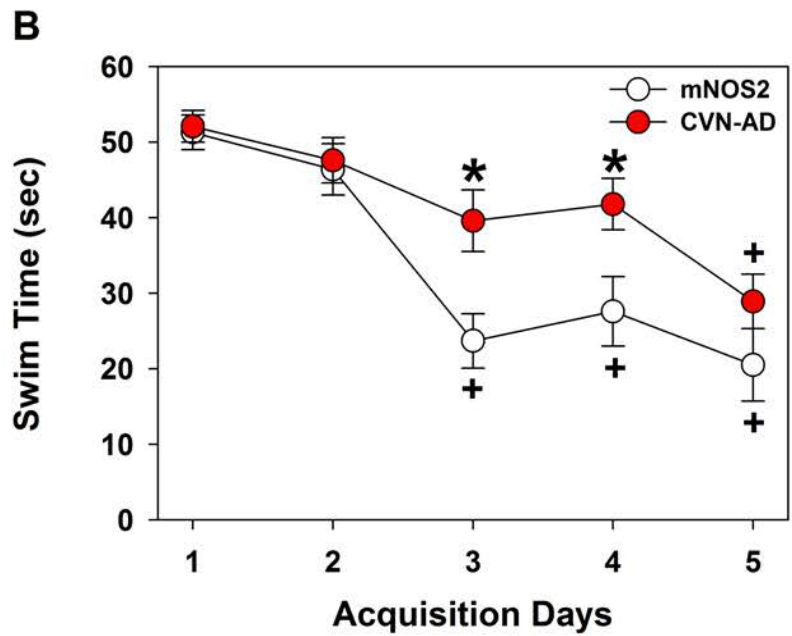
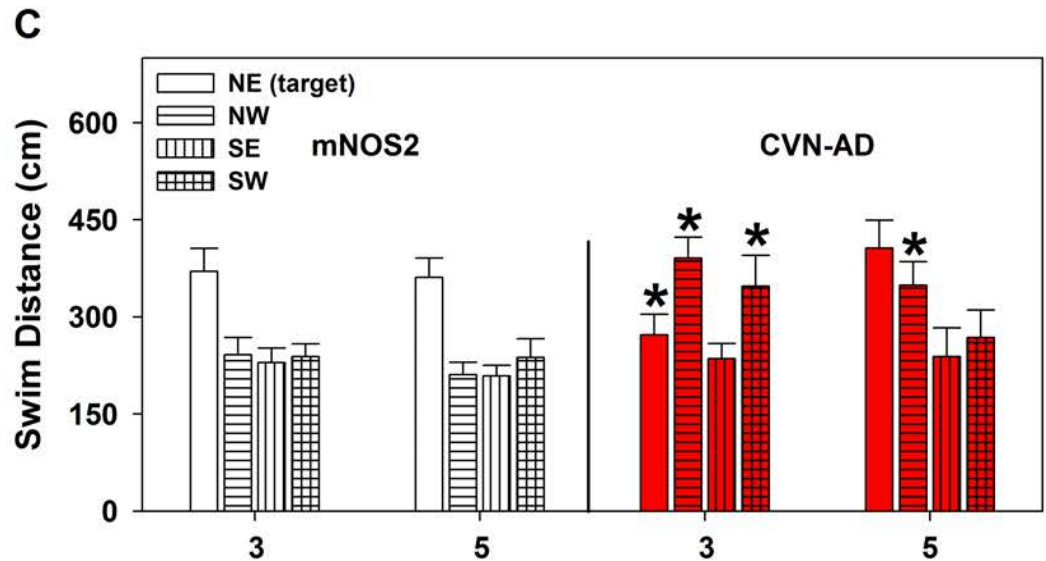
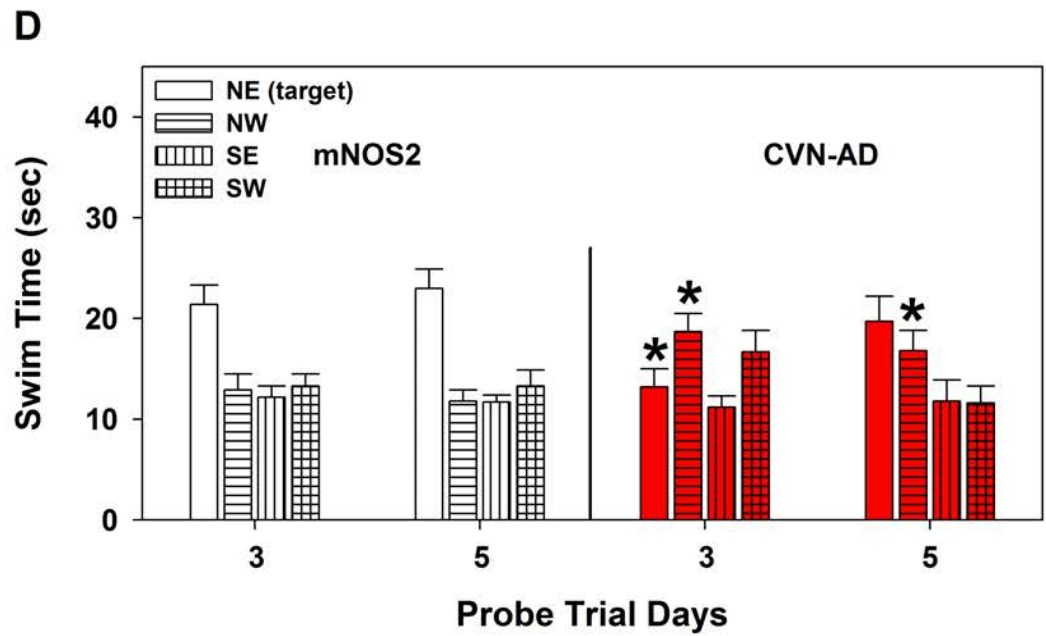

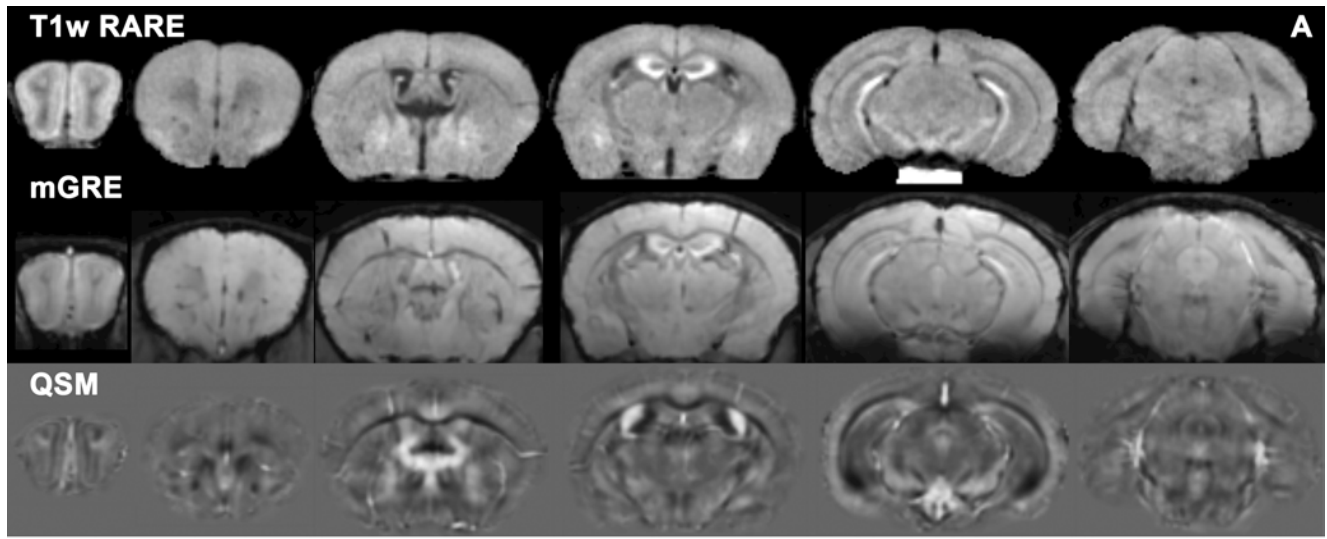
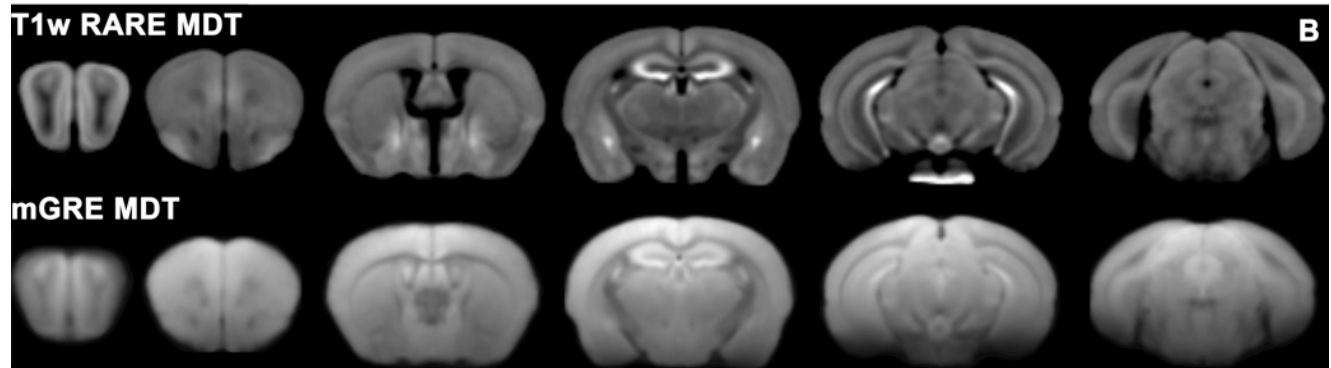
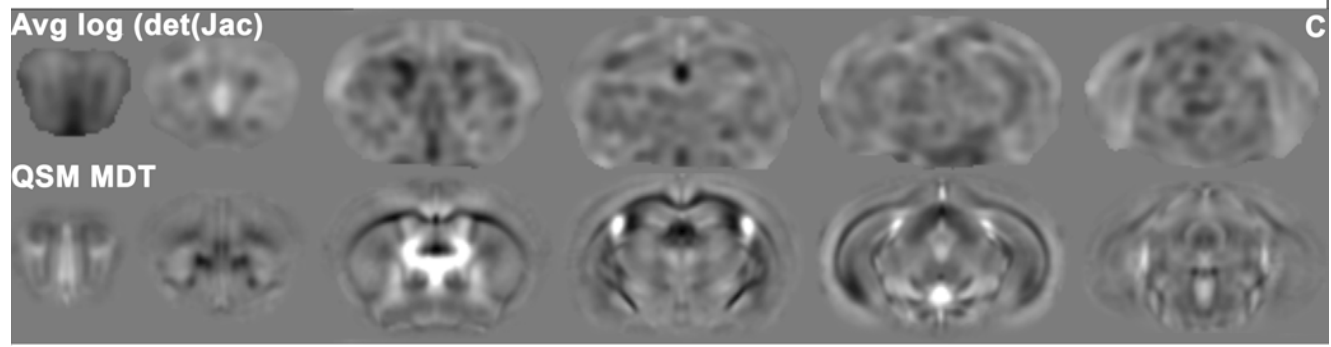

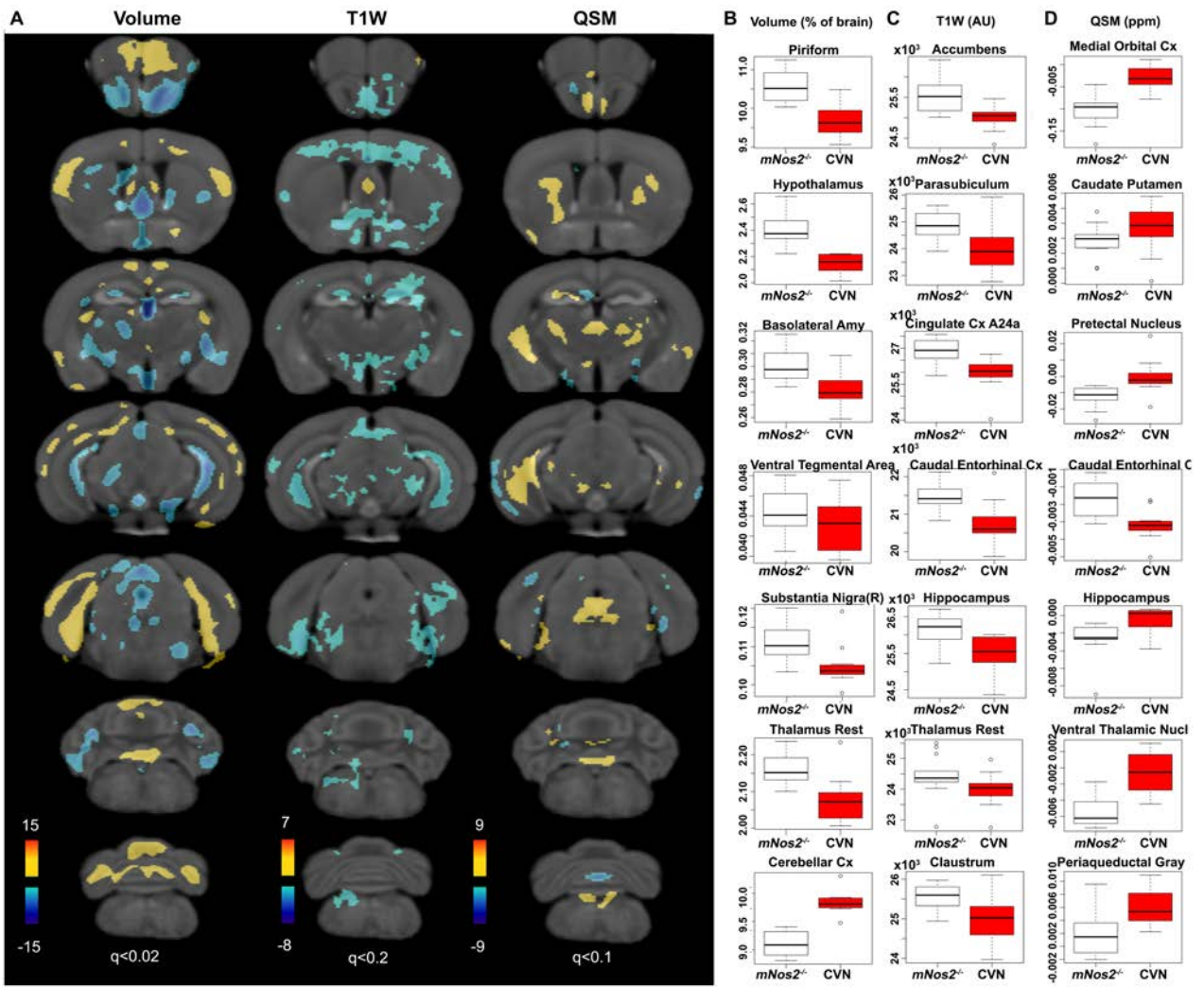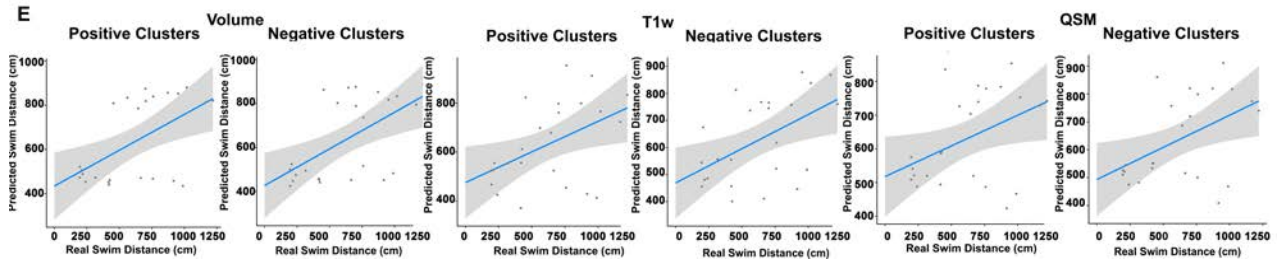

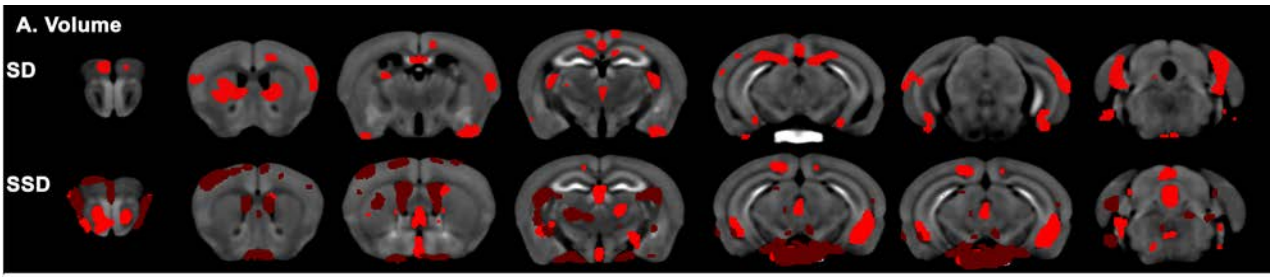
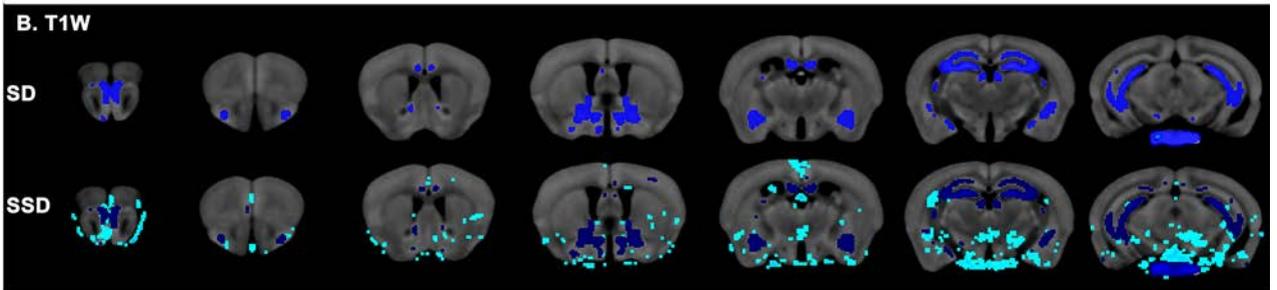
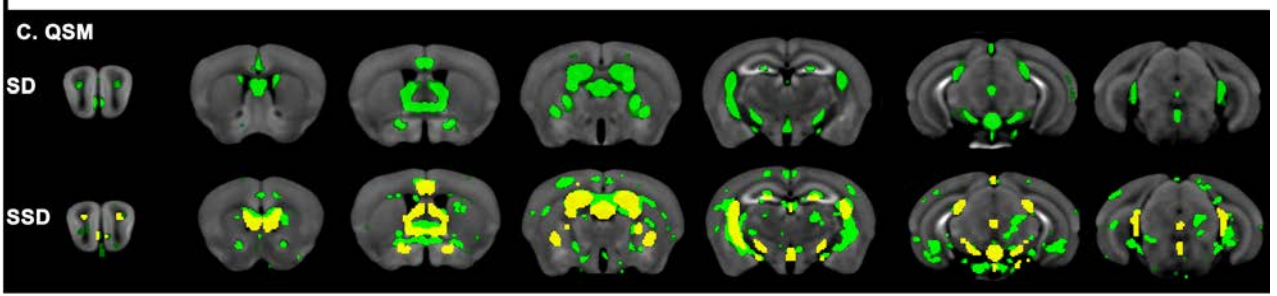
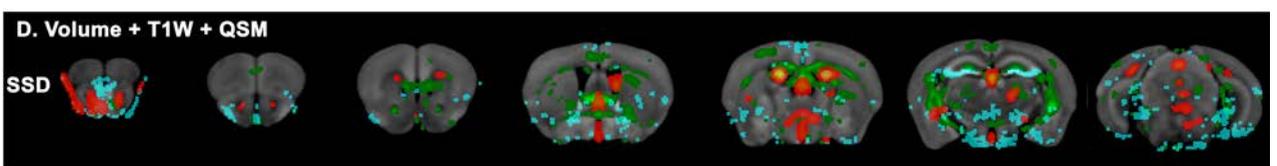
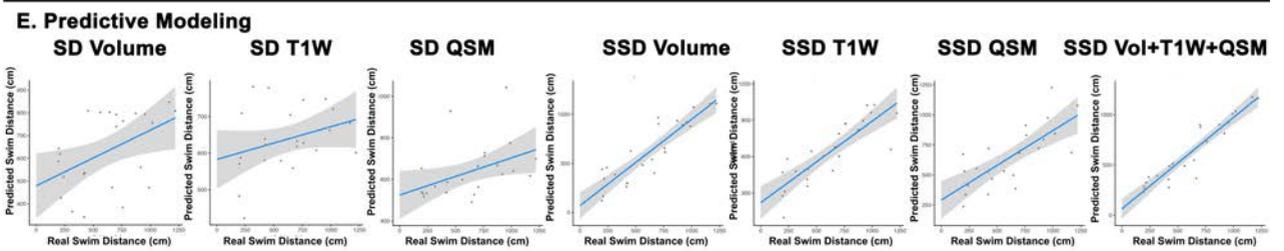

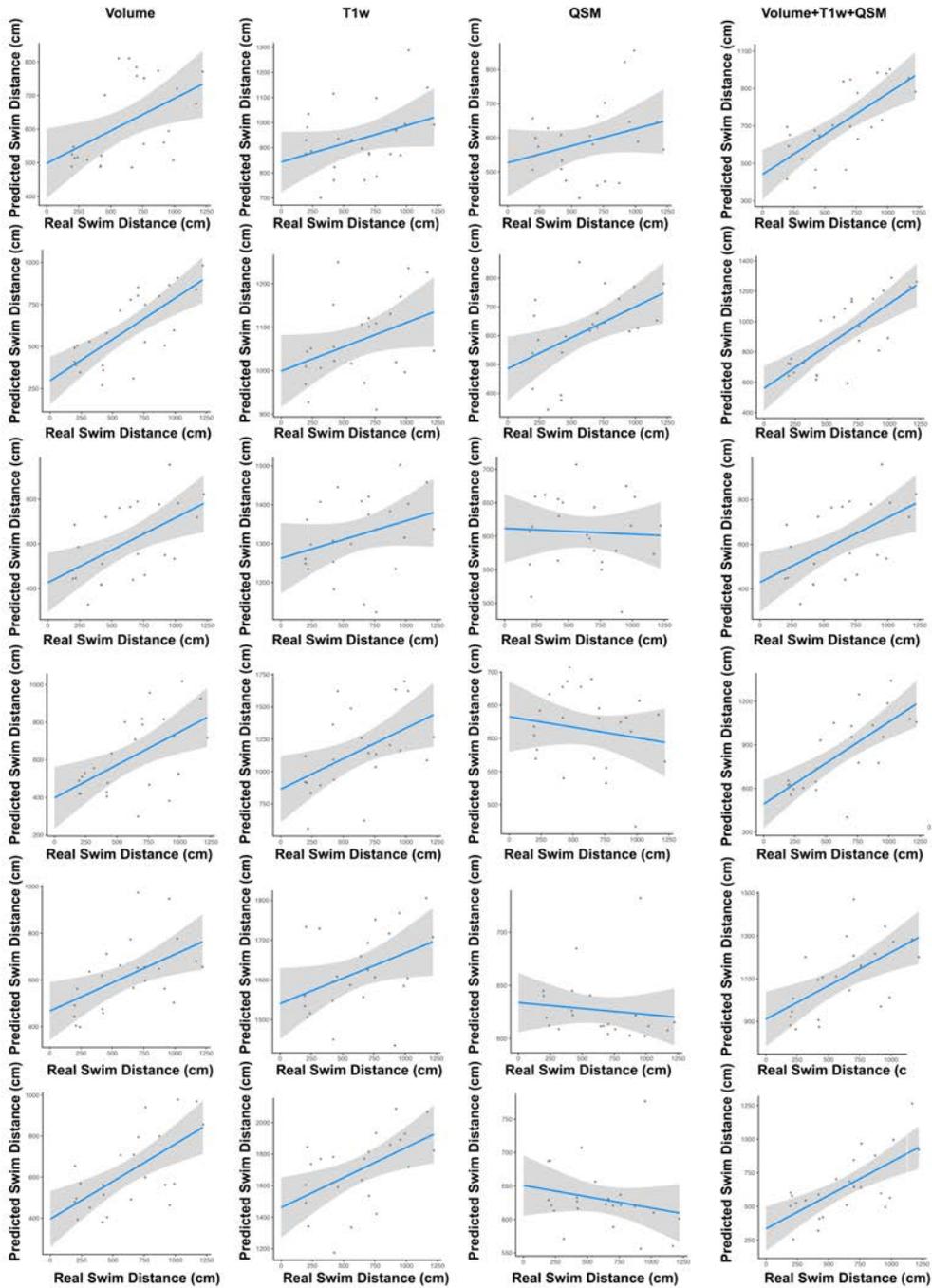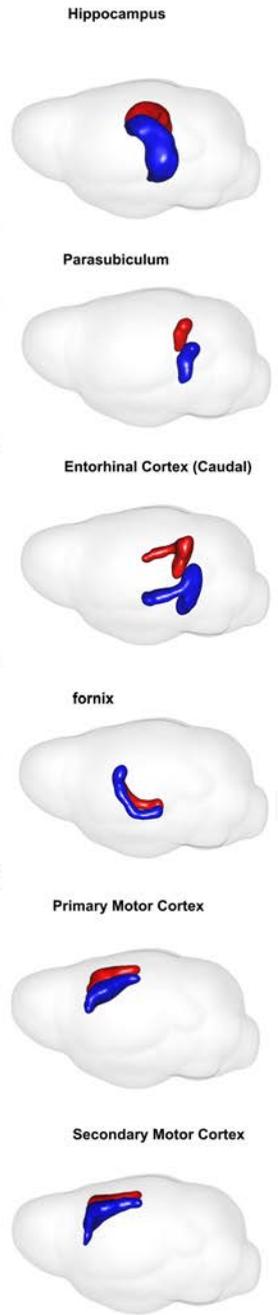

# Tables

**Table 1.** Model performance comparison based on a whole brain unbiased analysis, showing the testing root mean square error of the predictions (RMSE); and the whole set based Pearson correlation (corr), associated p value, and the explained variance (adjusted $R^2$).

| Contrast | Model | Test RMSE test (cm) | P | CORR | adjR2 | rank |
|---|---|---|---|---|---|---|
| VBA Volume | Positive Clusters | 266.94±56.84 | 0.005 | 0.55 | 0.27 | 5 |
| | Negative Clusters | 259.33±57.20 | 0.004 | 0.57 | 0.3 | 3 |
| | Combined Clusters | 261.76±58.31 | 0.004 | 0.57 | 0.3 | 4 |
| VBA T1W | Positive Clusters | 281.76±60.80 | 0.022 | 0.47 | 0.18 | 10 |
| | Negative Clusters | 273.98±54.30 | 0.01 | 0.51 | 0.23 | 7 |
| | Combined Clusters | 272.38±57.80 | 0.009 | 0.52 | 0.24 | 6 |
| VBA QSM | Positive Clusters | 285.55±73.88 | 0.036 | 0.43 | 0.15 | 11 |
| | Negative Clusters | 280.1±69.59 | 0.018 | 0.48 | 0.2 | 9 |
| | Combined Clusters | 290.6±50.53 | 0.01 | 0.52 | 0.23 | 12 |
| Sparse Decomposition (SD) | Volume | 300.7±48.89 | 0.021 | 0.47 | 0.18 | 14 |
| | T1W | 344.54±41.26 | 0.12 | 0.33 | 0.07 | 15 |
| | QSM | 299.19±79.46 | 0.04 | 0.43 | 0.15 | 13 |
| Supervised Sparse Decomposition (SD2) | Volume | 251.05±68.81 | 3.43E-09 | 0.9 | 0.79 | 2 |
| | T1W | 349.01±47.33 | 2.64E-07 | 0.84 | 0.69 | 16 |
| | QSM | 279±43.42 | 2.56E-05 | 0.75 | 0.54 | 8 |
| Multivariate SD2 | Volume+T1w+QSM | 233.08±101.78 | 7.14E-12 | 0.94 | 0.88 | 1 |

**Table 2.** To evaluate the 16 models performance, we have compared predictions with the true values for the measured behavior parameters (measured in cm) based on the full data set. Our Kruskall–Wallis analyses were followed by posthoc Tukey Kramer tests to control for the family wise error rate. Our results indicate that the supervised multivariate approach outperforms 12 of the other models. CI: confidence interval. VBA: voxel based analysis; SD2: supervised sparse decomposition.

| Model1 | Model2 | CI1 (cm) | Difference (cm) | CI2 (cm) | p |
|---|---|---|---|---|---|
| T1w_VBA- | COMBO_SD2 | 23.31 | 133.08 | 242.86 | 3.41E-03 |
| COMBO_SD2 | JAC_SD | -237.98 | -128.21 | -18.43 | 0.01 |
| T1w_VBA+ | COMBO_SD2 | 16.47 | 126.25 | 236.03 | 0.01 |
| COMBO_SD2 | JAC_VBA+ | -235.78 | -126.00 | -16.22 | 0.01 |
| T1w_VBA_COMBO | COMBO_SD2 | 14.72 | 124.50 | 234.28 | 0.01 |
| QSM_VBA_COMBO | COMBO_SD2 | 14.72 | 124.50 | 234.28 | 0.01 |
| COMBO_SD2 | JAC_VBA- | -231.48 | -121.71 | -11.93 | 0.01 |
| COMBO_SD2 | T1w_SD | -229.15 | -119.38 | -9.60 | 0.02 |
| COMBO_SD2 | JAC_VBA_COMBO | -227.65 | -117.88 | -8.10 | 0.02 |
| COMBO_SD2 | JAC_SD2 | -224.15 | -114.38 | -4.60 | 0.03 |
| COMBO_SD2 | QSM_SD2 | -223.57 | -113.79 | -4.02 | 0.03 |
| COMBO_SD2 | T1w_SD2 | -222.03 | -112.25 | -2.47 | 0.04 |

**Table 3.** Model performance comparison based on a ROI prior initialized based analysis, showing the testing root mean square error of the predictions (RMSE); and the whole set based Pearson correlation (corr), associated p value, and the explained variance (adjusted R2). Hc: hippocampus; PaS : parasubiculum; Cent: caudal entorhinal cortex; fx: fornix; M1: primary motor cortex; M2: secondary motor cortex.

| Contrast | Model | Test RMSE test (cm) | P | CORR | adjR2 | rank |
|---|---|---|---|---|---|---|
| Hc | volume | 288.57±93.684 | 0.013 | 0.49949 | 0.21538 | 7 |
| | T1 | 332.1±38.627 | 0.0968 | 0.34689 | 0.080345 | 17 |
| | QSM | 332.51±34.313 | 0.1588 | 0.29697 | 0.046743 | 18 |
| | combo | 285.59±82.017 | 9.86E-05 | 0.71095 | 0.48297 | 6 |
| PaS | volume | 239.09±37.498 | 5.16E-05 | 0.7299 | 0.51152 | 1 |
| | T1 | 334.78±27.95 | 0.0649 | 0.3827 | 0.10766 | 19 |
| | QSM | 283.29±90.102 | 0.0099 | 0.5158 | 0.23269 | 5 |
| | combo | 253.18±65.687 | 1.79E-05 | 0.75784 | 0.55498 | 2 |
| CEnt | volume | 268.77±60.513 | 0.0044 | 0.56087 | 0.28342 | 3 |
| | T1 | 342.14±86.897 | 0.1402 | 0.31018 | 0.055132 | 21 |
| | QSM | 325.09±72.639 | 0.8108 | 0.051575 | 0.042674 | 16 |
| | combo | 322.26±108.69 | 0.0044 | 0.56051 | 0.283 | 12 |
| fx | volume | 293.66±78.197 | 0.0062 | 0.54258 | 0.26232 | 8 |
| | T1 | 350.61±103.41 | 0.1589 | 0.29689 | 0.046694 | 23 |
| | QSM | 339.23±69.64 | 0.0314 | 0.4401 | 0.15704 | 20 |
| | combo | 323.41±120.91 | 0.2297 | 0.25469 | 0.022363 | 13 |
| M1 | volume | 323.83±131.29 | 0.0093 | 0.51959 | 0.2368 | 14 |
| | T1 | 312.18±113.8 | 0.0501 | 0.40417 | 0.12532 | 10 |
| | QSM | 360.89±94.909 | 0.5723 | 0.12131 | 0.030069 | 24 |
| | combo | 317.01±162.54 | 0.0019 | 0.60205 | 0.33348 | 11 |
| M2 | volume | 309.99±113.66 | 0.0008 | 0.63947 | 0.38205 | 9 |
| | T1 | 276.51±56.139 | 0.0085 | 0.52486 | 0.24254 | 4 |
| | QSM | 342.94±87.027 | 0.2891 | 0.22565 | 0.0077793 | 22 |
| | combo | 324.5±136 | 0.0003 | 0.67904 | 0.4366 | 15 |

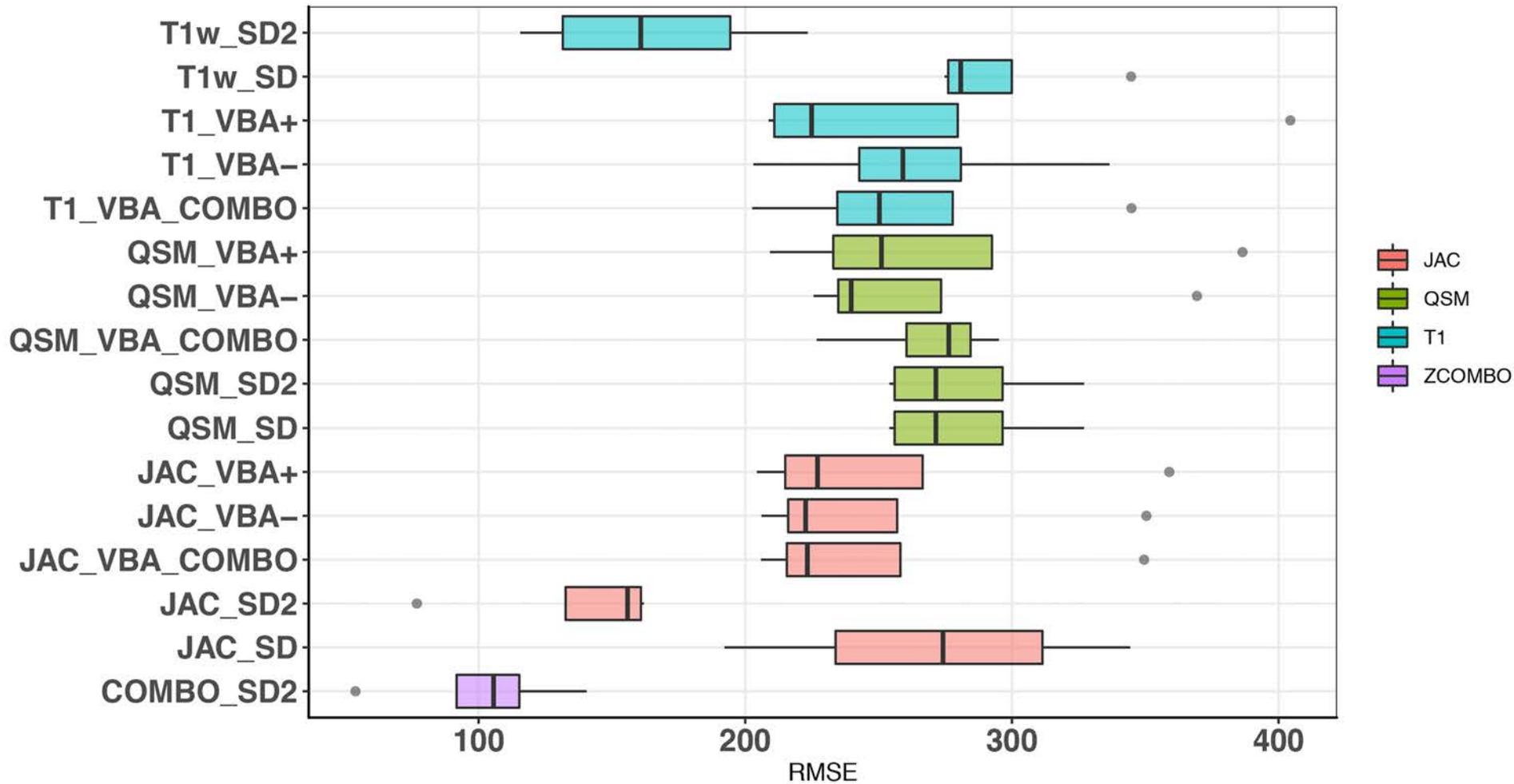

**Supplementary Table 1 Volume Differences**
(Ctrl: *mNos2-/-*; Abbr: Abbreviation; p BH: corrected p value using Benjamini Hochberg procedure, Diff: difference)

A. Atrophy

| Structure | Mean CVN-AD (%) | Mean Ctrl (%) | Std AD (%) | Std Ctrl (%) | p | p BH | CI1 (%) | CI2 (%) | t | Cohen d | Diff (%) | Abbr | Hemi-sphere |
|---|---|---|---|---|---|---|---|---|---|---|---|---|---|
| Anterior_Commisure | 0.13 | 0.14 | 0.0065 | 0.0039 | 0.0000 | 0.0001 | -0.0178 | -0.0089 | -6.23 | -2.55 | -9.24 | ac | Right |
| Anterior_Commisure | 0.14 | 0.15 | 0.0081 | 0.0063 | 0.0045 | 0.0183 | -0.0154 | -0.0032 | -3.16 | -1.29 | -6.31 | ac | Left |
| Basal Lateral Amygdala | 0.14 | 0.14 | 0.0089 | 0.0067 | 0.0067 | 0.0252 | -0.0161 | -0.0029 | -2.99 | -1.22 | -6.58 | BLA | Left |
| Basal Lateral Amygdala | 0.15 | 0.16 | 0.0073 | 0.0086 | 0.0061 | 0.0232 | -0.0168 | -0.0032 | -3.04 | -1.24 | -6.40 | BLA | Right |
| Bed_Nucleus_of_the_Stria_Terminalis | 0.09 | 0.09 | 0.0037 | 0.0033 | 0.0000 | 0.0002 | -0.0110 | -0.0051 | -5.68 | -2.33 | -8.55 | BNst | Left |
| Bed_Nucleus_of_the_Stria_Terminalis | 0.08 | 0.09 | 0.0073 | 0.0036 | 0.0043 | 0.0177 | -0.0120 | -0.0025 | -3.18 | -1.30 | -7.98 | BNst | Right |
| Cerebral_Peduncle | 0.19 | 0.21 | 0.0073 | 0.0091 | 0.0001 | 0.0008 | -0.0237 | -0.0095 | -4.84 | -1.98 | -7.93 | cp | Right |
| Cerebral_Peduncle | 0.21 | 0.22 | 0.0042 | 0.0078 | 0.0002 | 0.0016 | -0.0171 | -0.0062 | -4.42 | -1.81 | -5.28 | cp | Left |
| Deep_Mesencephalic_Nuclei | 0.25 | 0.27 | 0.0168 | 0.0085 | 0.0001 | 0.0009 | -0.0363 | -0.0143 | -4.76 | -1.95 | -9.22 | DpMe | Left |
| Dorsal_Tegmentum | 0.18 | 0.2 | 0.0079 | 0.0050 | 0.0000 | 0.0000 | -0.0239 | -0.0129 | -6.95 | -2.84 | -9.39 | NA | Right |
| Dorsal_Tegmentum | 0.18 | 0.2 | 0.0064 | 0.0053 | 0.0001 | 0.0007 | -0.0165 | -0.0067 | -4.87 | -1.99 | -5.92 | NA | Left |
| Fastigial_Medial_Dorsolateral_Nucleus_of_Cerebellum | 0.01 | 0.01 | 0.0011 | 0.0008 | 0.0119 | 0.0376 | -0.0018 | -0.0002 | -2.74 | -1.12 | -10.10 | FasDL | Left |
| Fastigial_Medial_Dorsolateral_Nucleus_of_Cerebellum | 0.01 | 0.01 | 0.0009 | 0.0007 | 0.0116 | 0.0372 | -0.0016 | -0.0002 | -2.75 | -1.13 | -8.64 | FasDL | Right |
| Fimbria | 0.23 | 0.25 | 0.0100 | 0.0107 | 0.0056 | 0.0221 | -0.0219 | -0.0042 | -3.07 | -1.26 | -5.32 | fi | Left |
| Fornix | 0.04 | 0.05 | 0.0024 | 0.0029 | 0.0000 | 0.0000 | -0.0099 | -0.0054 | -7.05 | -2.89 | -15.30 | fx | Left |
| Fornix | 0.05 | 0.05 | 0.0017 | 0.0025 | 0.0000 | 0.0000 | -0.0088 | -0.0051 | -7.88 | -3.23 | -12.80 | fx | Right |
| Hypothalamus | 1.08 | 1.24 | 0.0468 | 0.0835 | 0.0000 | 0.0003 | -0.2100 | -0.0928 | -5.34 | -2.19 | -12.30 | Hyp | Right |
| Hypothalamus | 1.07 | 1.18 | 0.0386 | 0.0442 | 0.0000 | 0.0000 | -0.1460 | -0.0753 | -6.48 | -2.66 | -9.42 | Hyp | Left |
| Internal_Capsule | 0.41 | 0.44 | 0.0132 | 0.0158 | 0.0006 | 0.0040 | -0.0365 | -0.0116 | -4.00 | -1.64 | -5.47 | ic | Left |
| Internal_Capsule | 0.39 | 0.41 | 0.0147 | 0.0200 | 0.0071 | 0.0259 | -0.0367 | -0.0065 | -2.97 | -1.22 | -5.28 | ic | Right |
| Interpeduncular_Nucleus | 0.03 | 0.03 | 0.0025 | 0.0022 | 0.0186 | 0.0509 | -0.0044 | -0.0004 | -2.54 | -1.04 | -7.94 | Iped | Right |
| Lateral_Olfactory_Tract | 0.16 | 0.18 | 0.0091 | 0.0291 | 0.0099 | 0.0329 | -0.0449 | -0.0069 | -2.82 | -1.16 | -14.00 | lo | Right |
| Mamillothalamic_Tract | 0.01 | 0.01 | 0.0013 | 0.0012 | 0.0074 | 0.0265 | -0.0025 | -0.0004 | -2.95 | -1.21 | -10.30 | mt | Right |
| Optic_Tracts | 0.25 | 0.27 | 0.0181 | 0.0257 | 0.0095 | 0.0324 | -0.0455 | -0.0071 | -2.84 | -1.17 | -9.58 | ot | Left |
| Optic_Tracts | 0.26 | 0.27 | 0.0151 | 0.0178 | 0.0152 | 0.0433 | -0.0321 | -0.0038 | -2.63 | -1.08 | -6.56 | ot | Right |

| Region | | | | | | | | | | | | Abbr | Side |
|---|---|---|---|---|---|---|---|---|---|---|---|---|---|
| Periaquaductal_Grey | 0.51 | 0.53 | 0.0234 | 0.0122 | 0.0196 | 0.0521 | -0.0342 | -0.0033 | -2.52 | -1.03 | -3.54 | PAG | Right |
| Piriform_Cortex | 5.06 | 5.54 | 0.2300 | 0.2040 | 0.0000 | 0.0003 | -0.6600 | -0.2920 | -5.37 | -2.20 | -8.60 | Pir | Right |
| Piriform_Cortex | 5.13 | 5.5 | 0.2140 | 0.2390 | 0.0007 | 0.0047 | -0.5600 | -0.1730 | -3.92 | -1.61 | -6.66 | Pir | Left |
| Pontine_Reticular_Nucleus | 0.61 | 0.65 | 0.0310 | 0.0283 | 0.0124 | 0.0385 | -0.0581 | -0.0079 | -2.72 | -1.12 | -5.10 | PnC | Left |
| Pontine_Reticular_Nucleus | 0.56 | 0.59 | 0.0174 | 0.0133 | 0.0001 | 0.0009 | -0.0426 | -0.0167 | -4.74 | -1.94 | -5.03 | PnC | Right |
| PosteriorDorsal_ParaventricularMedialParvicellular_Posterior_LateralHypothalamus | 0.12 | 0.12 | 0.0066 | 0.0056 | 0.0054 | 0.0216 | -0.0128 | -0.0025 | -3.09 | -1.26 | -6.19 | PHD_LatHy | Right |
| PosteriorDorsal_ParaventricularMedialParvicellular_Posterior_LateralHypothalamus | 0.11 | 0.12 | 0.0040 | 0.0058 | 0.0040 | 0.0165 | -0.0110 | -0.0024 | -3.22 | -1.32 | -5.73 | PHD_LatHy | Left |
| Retro_Rubral_Field | 0.04 | 0.05 | 0.0018 | 0.0023 | 0.0125 | 0.0385 | -0.0041 | -0.0006 | -2.72 | -1.12 | -5.14 | RR | Left |
| Rostral_Linear_Nucleus | 0.03 | 0.04 | 0.0037 | 0.0026 | 0.0172 | 0.0472 | -0.0060 | -0.0007 | -2.58 | -1.05 | -9.07 | RI | Right |
| Stria_Terminalis | 0.09 | 0.1 | 0.0036 | 0.0041 | 0.0001 | 0.0012 | -0.0106 | -0.0040 | -4.60 | -1.89 | -7.19 | st | Left |
| Striatum | 2.2 | 2.28 | 0.0720 | 0.0827 | 0.0203 | 0.0534 | -0.1460 | -0.0136 | -2.50 | -1.03 | -3.50 | Cpu | Left |
| Substantia_Nigra | 0.11 | 0.11 | 0.0055 | 0.0050 | 0.0137 | 0.0408 | -0.0102 | -0.0013 | -2.68 | -1.10 | -5.18 | SN | Right |
| Subthalamic_Nucleus | 0.05 | 0.05 | 0.0018 | 0.0019 | 0.0000 | 0.0001 | -0.0064 | -0.0032 | -6.25 | -2.56 | -9.34 | Sthal | Left |
| Subthalamic_Nucleus | 0.05 | 0.05 | 0.0034 | 0.0030 | 0.0039 | 0.0164 | -0.0070 | -0.0015 | -3.22 | -1.32 | -8.32 | Sthal | Right |
| Superior_Cerebellar_Peduncle | 0.09 | 0.1 | 0.0043 | 0.0054 | 0.0002 | 0.0016 | -0.0131 | -0.0047 | -4.41 | -1.81 | -8.88 | scp | Right |
| Superior_Cerebellar_Peduncle | 0.1 | 0.1 | 0.0057 | 0.0047 | 0.0062 | 0.0233 | -0.0108 | -0.0020 | -3.03 | -1.24 | -6.27 | scp | Left |
| Superior_Colliculus | 1.06 | 1.12 | 0.0464 | 0.0362 | 0.0021 | 0.0101 | -0.0937 | -0.0238 | -3.49 | -1.43 | -5.27 | SC | Right |
| Superior_Colliculus | 1.04 | 1.09 | 0.0412 | 0.0270 | 0.0009 | 0.0053 | -0.0828 | -0.0247 | -3.84 | -1.57 | -4.93 | SC | Left |
| Tegmental_Nucleus | 0 | 0 | 0.0005 | 0.0004 | 0.0142 | 0.0411 | -0.0008 | -0.0001 | -2.66 | -1.09 | -9.96 | Tg | Right |
| Thalamus_Rest | 1.05 | 1.1 | 0.0371 | 0.0266 | 0.0027 | 0.0128 | -0.0710 | -0.0170 | -3.38 | -1.38 | -4.02 | NA | Right |
| Thalamus_Rest | 1.03 | 1.07 | 0.0334 | 0.0226 | 0.0022 | 0.0106 | -0.0635 | -0.0159 | -3.46 | -1.42 | -3.73 | NA | Left |
| Zona_Incerta | 0.22 | 0.23 | 0.0091 | 0.0082 | 0.0009 | 0.0053 | -0.0209 | -0.0063 | -3.84 | -1.57 | -5.87 | ZI | Right |
| Zona_Incerta | 0.24 | 0.25 | 0.0094 | 0.0072 | 0.0029 | 0.0133 | -0.0185 | -0.0044 | -3.35 | -1.37 | -4.55 | ZI | Left |

B. Hypertrophy

(Ctrl: *mNos2*-/-; Abbr: Abbreviation; p BH: corrected p value using Benjamini Hochberg procedure, Diff: difference)

| Structure | Mean CVN-AD (%) | Mean Ctrl (%) | Std AD (%) | Std Ctrl *%) | p | p BH | CI1 (%) | CI2 (%) | t | Cohen d | Diff (%) | Abbr | Hemi-sphere |
|---|---|---|---|---|---|---|---|---|---|---|---|---|---|
| Caudomedial_Entorhinal_Cortex | 0.599 | 0.559 | 0.021 | 0.015 | 0.0000 | 0.0002 | 0.025 | 0.055 | 5.507 | 2.253 | 7.175 | CEnt | Left |
| Caudomedial_Entorhinal_Cortex | 0.573 | 0.534 | 0.017 | 0.019 | 0.0000 | 0.0004 | 0.023 | 0.054 | 5.205 | 2.133 | 7.267 | CEnt | Right |
| Cerebellar_Cortex | 4.985 | 4.644 | 0.105 | 0.154 | 0.0000 | 0.0001 | 0.228 | 0.455 | 6.232 | 2.557 | 7.352 | 1Cb_to_10Cb | Left |
| Cerebellar_Cortex | 4.85 | 4.458 | 0.108 | 0.117 | 0.0000 | 0.0000 | 0.296 | 0.488 | 8.464 | 3.468 | 8.792 | 1Cb_to_10Cb | Right |
| Cerebellar_White_Matter | 1.183 | 1.127 | 0.021 | 0.032 | 0.0001 | 0.0007 | 0.032 | 0.079 | 4.908 | 2.014 | 4.903 | cbw | Left |
| Cerebellar_White_Matter | 1.215 | 1.151 | 0.022 | 0.03 | 0.0000 | 0.0001 | 0.041 | 0.087 | 5.863 | 2.404 | 5.566 | cbw | Right |
| Cingulate_Cortex_Area_24a | 0.195 | 0.18 | 0.018 | 0.01 | 0.0147 | 0.0422 | 0.003 | 0.027 | 2.648 | 1.082 | 8.523 | A24a | Right |
| Cingulate_Cortex_Area_24b | 0.165 | 0.15 | 0.014 | 0.006 | 0.0015 | 0.0084 | 0.006 | 0.023 | 3.617 | 1.477 | 9.928 | A24b | Right |
| Cingulate_Cortex_Area_24b_prime | 0.046 | 0.042 | 0.003 | 0.003 | 0.0013 | 0.0075 | 0.002 | 0.006 | 3.685 | 1.509 | 9.231 | A24bPrime | Left |
| Cingulate_Cortex_Area_24b_prime | 0.055 | 0.047 | 0.005 | 0.002 | 0.0000 | 0.0003 | 0.005 | 0.01 | 5.391 | 2.202 | 15.562 | A24bPrime | Right |
| Cingulate_Cortex_Area_29a | 0.055 | 0.049 | 0.005 | 0.003 | 0.0031 | 0.0139 | 0.002 | 0.01 | 3.329 | 1.362 | 12.109 | A29a | Left |
| Cingulate_Cortex_Area_29c | 0.14 | 0.13 | 0.013 | 0.006 | 0.0165 | 0.0457 | 0.002 | 0.019 | 2.597 | 1.061 | 7.921 | A29c | Right |
| Cingulate_Cortex_Area_30 | 0.292 | 0.269 | 0.009 | 0.009 | 0.0000 | 0.0001 | 0.015 | 0.03 | 6.31 | 2.585 | 8.470 | A30 | Left |
| Cingulate_Cortex_Area_30 | 0.298 | 0.272 | 0.01 | 0.013 | 0.0000 | 0.0003 | 0.016 | 0.036 | 5.354 | 2.196 | 9.542 | A30 | Right |
| Cingulate_Cortex_Area_32 | 0.22 | 0.2 | 0.013 | 0.011 | 0.0009 | 0.0053 | 0.009 | 0.029 | 3.845 | 1.574 | 9.514 | A32 | Right |
| Claustrum | 0.032 | 0.028 | 0.003 | 0.002 | 0.0034 | 0.0149 | 0.001 | 0.006 | 3.281 | 1.342 | 12.442 | Cl | Right |
| Dorsal_Claustrum | 0.013 | 0.011 | 0.001 | 0.001 | 0.0015 | 0.0084 | 0.001 | 0.002 | 3.617 | 1.481 | 12.247 | DCl | Right |
| Dorsal_Claustrum | 0.017 | 0.015 | 0.002 | 0.002 | 0.0018 | 0.0094 | 0.001 | 0.004 | 3.547 | 1.452 | 16.962 | DCl | Left |
| Dorsal_Tenia_Tecta | 0.051 | 0.046 | 0.004 | 0.003 | 0.0038 | 0.0161 | 0.002 | 0.008 | 3.239 | 1.325 | 10.965 | DTT | Left |
| Dorsal_Tenia_Tecta | 0.056 | 0.049 | 0.005 | 0.006 | 0.0080 | 0.0283 | 0.002 | 0.011 | 2.917 | 1.195 | 13.324 | DTT | Right |
| Ectorhinal_Cortex | 0.263 | 0.253 | 0.009 | 0.01 | 0.0162 | 0.0454 | 0.002 | 0.018 | 2.604 | 1.067 | 3.925 | Ect | Right |
| Ectorhinal_Cortex | 0.286 | 0.261 | 0.007 | 0.015 | 0.0000 | 0.0005 | 0.015 | 0.035 | 5.121 | 2.103 | 9.554 | Ect | Left |
| Frontal_Association_Cortex | 0.88 | 0.83 | 0.049 | 0.038 | 0.0103 | 0.0334 | 0.013 | 0.086 | 2.805 | 1.148 | 5.986 | FrA | Right |
| Frontal_Cortex_Area_3 | 0.096 | 0.09 | 0.006 | 0.004 | 0.0068 | 0.0253 | 0.002 | 0.01 | 2.984 | 1.221 | 6.723 | Fr3 | Left |

| Region | | | | | | | | | | | | | |
|---|---|---|---|---|---|---|---|---|---|---|---|---|---|
| Frontal_Cortex_Area_3 | 0.102 | 0.092 | 0.007 | 0.004 | 0.0002 | 0.0016 | 0.005 | 0.015 | 4.402 | 1.799 | 10.899 | Fr3 | Right |
| Insular_Cortex | 0.705 | 0.676 | 0.036 | 0.019 | 0.0189 | 0.0511 | 0.005 | 0.053 | 2.535 | 1.036 | 4.316 | Ins | Right |
| Insular_Cortex | 0.797 | 0.738 | 0.027 | 0.03 | 0.0001 | 0.0006 | 0.034 | 0.083 | 4.984 | 2.043 | 7.980 | Ins | Left |
| Lateral_Orbital_Cortex | 0.365 | 0.327 | 0.027 | 0.011 | 0.0001 | 0.0011 | 0.021 | 0.055 | 4.625 | 1.889 | 11.681 | LO | Right |
| Latero_Posterior_Nuclei_of_Thalamus | 0.018 | 0.016 | 0.002 | 0.001 | 0.0157 | 0.0442 | 0 | 0.003 | 2.62 | 1.071 | 9.566 | LP | Left |
| Medial_Entorhinal_Cortex | 0.086 | 0.079 | 0.008 | 0.005 | 0.0101 | 0.0332 | 0.002 | 0.013 | 2.816 | 1.152 | 9.397 | MEnt | Right |
| Medial_Entorhinal_Cortex | 0.089 | 0.08 | 0.008 | 0.004 | 0.0019 | 0.0096 | 0.003 | 0.013 | 3.523 | 1.44 | 10.456 | MEnt | Left |
| Medial_Geniculate_Nucleus | 0.043 | 0.041 | 0.002 | 0.002 | 0.0017 | 0.0092 | 0.001 | 0.004 | 3.564 | 1.461 | 6.308 | MGN | Left |
| Medial_Orbital_Cortex | 0.14 | 0.121 | 0.01 | 0.008 | 0.0001 | 0.0006 | 0.011 | 0.026 | 5.017 | 2.054 | 15.139 | MO | Left |
| Medial_Orbital_Cortex | 0.151 | 0.126 | 0.007 | 0.007 | 0.0000 | 0.0000 | 0.019 | 0.031 | 8.831 | 3.618 | 19.789 | MO | Right |
| Medial_Parietal_Association_Cortex | 0.035 | 0.032 | 0.003 | 0.002 | 0.0134 | 0.0405 | 0.001 | 0.005 | 2.69 | 1.1 | 7.989 | MPtA | Right |
| Medial_Parietal_Association_Cortex | 0.036 | 0.033 | 0.003 | 0.002 | 0.0133 | 0.0405 | 0.001 | 0.005 | 2.693 | 1.101 | 8.914 | MPtA | Left |
| Parasubiculum | 0.106 | 0.087 | 0.006 | 0.005 | 0.0000 | 0.0000 | 0.014 | 0.023 | 8.611 | 3.526 | 21.189 | PaS | Left |
| Parasubiculum | 0.125 | 0.101 | 0.007 | 0.008 | 0.0000 | 0.0000 | 0.017 | 0.029 | 7.622 | 3.125 | 22.766 | PaS | Right |
| Perirhinal_Cortex | 0.224 | 0.214 | 0.006 | 0.011 | 0.0135 | 0.0405 | 0.002 | 0.017 | 2.686 | 1.102 | 4.523 | PRh | Left |
| Postsubiculum | 0.096 | 0.09 | 0.004 | 0.005 | 0.0033 | 0.0147 | 0.002 | 0.01 | 3.294 | 1.35 | 6.535 | Post | Left |
| Postsubiculum | 0.096 | 0.086 | 0.008 | 0.007 | 0.0056 | 0.0221 | 0.003 | 0.016 | 3.068 | 1.257 | 10.793 | Post | Right |
| Presubiculum | 0.017 | 0.015 | 0.001 | 0.001 | 0.0006 | 0.0038 | 0.001 | 0.003 | 4.028 | 1.652 | 13.215 | PrS | Right |
| Presubiculum | 0.019 | 0.016 | 0.002 | 0.001 | 0.0002 | 0.0016 | 0.001 | 0.004 | 4.412 | 1.805 | 16.854 | PrS | Left |
| Primary_Auditory_Cortex | 0.164 | 0.155 | 0.005 | 0.006 | 0.0008 | 0.0052 | 0.004 | 0.013 | 3.87 | 1.587 | 5.392 | Au1 | Right |
| Primary_Auditory_Cortex | 0.176 | 0.165 | 0.005 | 0.007 | 0.0003 | 0.0020 | 0.006 | 0.017 | 4.298 | 1.763 | 6.764 | Au1 | Left |
| Primary_Motor_Cortex | 0.71 | 0.672 | 0.03 | 0.021 | 0.0017 | 0.0089 | 0.016 | 0.059 | 3.584 | 1.466 | 5.583 | M1 | Right |
| Primary_Somatosensory_Cortex | 0.084 | 0.079 | 0.005 | 0.003 | 0.0038 | 0.0161 | 0.002 | 0.009 | 3.238 | 1.325 | 6.681 | S1 | Left |
| Primary_Somatosensory_Cortex | 0.094 | 0.087 | 0.006 | 0.005 | 0.0058 | 0.0224 | 0.002 | 0.012 | 3.057 | 1.251 | 8.361 | S1 | Right |
| Primary_Somatosensory_Cortex_Barrel_Field | 0.883 | 0.839 | 0.033 | 0.025 | 0.0015 | 0.0084 | 0.018 | 0.068 | 3.624 | 1.483 | 5.130 | S1BF | Left |
| Primary_Somatosensory_Cortex_Dysgranular_Zone | 0.135 | 0.126 | 0.01 | 0.003 | 0.0072 | 0.0259 | 0.003 | 0.014 | 2.965 | 1.211 | 6.732 | S1DZ | Right |
| Primary_Somatosensory_Cortex_Jaw_Region | 0.283 | 0.263 | 0.012 | 0.01 | 0.0002 | 0.0012 | 0.011 | 0.03 | 4.568 | 1.87 | 7.818 | S1J | Right |
| Primary_Somatosensory_Cortex_Jaw_Region | 0.305 | 0.279 | 0.017 | 0.012 | 0.0003 | 0.0019 | 0.014 | 0.039 | 4.326 | 1.77 | 9.350 | S1J | Left |
| Primary_Somatosensory_Cortex_Upper_Lip_Region | 0.416 | 0.399 | 0.015 | 0.014 | 0.0086 | 0.0297 | 0.005 | 0.03 | 2.887 | 1.182 | 4.348 | S1ULp | Right |
| Primary_Somatosensory_Cortex_Upper_Lip_Region | 0.432 | 0.391 | 0.018 | 0.012 | 0.0000 | 0.0000 | 0.028 | 0.054 | 6.497 | 2.658 | 10.512 | S1ULp | Left |
| Primary_Somatosensory_CortexForelimb_Region | 0.328 | 0.304 | 0.015 | 0.009 | 0.0001 | 0.0008 | 0.014 | 0.035 | 4.79 | 1.959 | 7.996 | S1FL | Right |
| Primary_Visual_Cortex_Binocular_Area | 0.192 | 0.178 | 0.016 | 0.011 | 0.0213 | 0.0553 | 0.002 | 0.025 | 2.48 | 1.015 | 7.700 | V1B | Left |
| Primary_Visual_Cortex_Binocular_Area | 0.196 | 0.182 | 0.006 | 0.007 | 0.0000 | 0.0004 | 0.009 | 0.02 | 5.269 | 2.159 | 8.081 | V1B | Right |
| Primary_Visual_Cortex_Monocular_Area | 0.451 | 0.429 | 0.024 | 0.011 | 0.0083 | 0.0292 | 0.006 | 0.038 | 2.898 | 1.184 | 5.115 | V1M | Right |
| Primary_Visual_Cortex_Monocular_Area | 0.415 | 0.391 | 0.026 | 0.018 | 0.0140 | 0.0411 | 0.005 | 0.043 | 2.668 | 1.092 | 6.214 | V1M | Left |
| Pyramidal_Tract | 0.1 | 0.089 | 0.012 | 0.008 | 0.0097 | 0.0328 | 0.003 | 0.019 | 2.83 | 1.158 | 12.458 | py | Right |
| Pyramidal_Tract | 0.109 | 0.095 | 0.014 | 0.011 | 0.0141 | 0.0411 | 0.003 | 0.024 | 2.665 | 1.091 | 14.023 | py | Left |
| Secondary_Auditory_Cortex_Dorsal_Part | 0.191 | 0.183 | 0.008 | 0.008 | 0.0204 | 0.0534 | 0.001 | 0.015 | 2.5 | 1.024 | 4.382 | AuD | Right |
| Secondary_Auditory_Cortex_Dorsal_Part | 0.198 | 0.188 | 0.01 | 0.007 | 0.0095 | 0.0324 | 0.003 | 0.017 | 2.84 | 1.162 | 5.229 | AuD | Left |
| Secondary_Auditory_Cortex_Ventral_Part | 0.201 | 0.193 | 0.004 | 0.009 | 0.0122 | 0.0382 | 0.002 | 0.014 | 2.732 | 1.122 | 4.150 | AuV | Left |
| Secondary_Motor_Cortex | 0.636 | 0.609 | 0.027 | 0.025 | 0.0194 | 0.0521 | 0.005 | 0.049 | 2.521 | 1.032 | 4.386 | M2 | Left |
| Secondary_Motor_Cortex | 0.68 | 0.63 | 0.033 | 0.019 | 0.0001 | 0.0010 | 0.028 | 0.073 | 4.698 | 1.921 | 8.052 | M2 | Right |
| Secondary_Somatosensory_Cortex | 0.68 | 0.632 | 0.02 | 0.018 | 0.0000 | 0.0001 | 0.032 | 0.064 | 6.091 | 2.494 | 7.575 | S2 | Right |
| Secondary_Somatosensory_Cortex | 0.646 | 0.58 | 0.025 | 0.024 | 0.0000 | 0.0000 | 0.045 | 0.087 | 6.56 | 2.687 | 11.382 | S2 | Left |
| Secondary_Visual_CortexLateral_Area | 0.27 | 0.253 | 0.012 | 0.008 | 0.0004 | 0.0030 | 0.008 | 0.025 | 4.125 | 1.688 | 6.489 | V2L | Right |
| Secondary_Visual_CortexLateral_Area | 0.277 | 0.252 | 0.013 | 0.014 | 0.0001 | 0.0012 | 0.014 | 0.036 | 4.585 | 1.879 | 9.923 | V2L | Left |
| Temporal_Association_Cortex | 0.326 | 0.304 | 0.013 | 0.008 | 0.0000 | 0.0004 | 0.013 | 0.031 | 5.141 | 2.102 | 7.304 | TeA | Right |
| Temporal_Association_Cortex | 0.343 | 0.311 | 0.011 | 0.014 | 0.0000 | 0.0001 | 0.021 | 0.043 | 6.158 | 2.525 | 10.299 | TeA | Left |
| Ventral_Intermediate_Entorhinal_Cortex | 0.148 | 0.135 | 0.01 | 0.008 | 0.0020 | 0.0097 | 0.005 | 0.02 | 3.508 | 1.436 | 9.346 | VIEnt | Right |
| Ventral_Orbital_Cortex | 0.15 | 0.139 | 0.012 | 0.007 | 0.0103 | 0.0334 | 0.003 | 0.02 | 2.805 | 1.147 | 8.201 | VO | Left |
| Ventral_Orbital_Cortex | 0.156 | 0.14 | 0.01 | 0.004 | 0.0000 | 0.0002 | 0.01 | 0.023 | 5.538 | 2.263 | 11.968 | VO | Right |

**Supplementary Table 2. T1w signal intensity differences**
(Ctrl: *mNos2-/-*; Abbr: Abbreviation; p BH: corrected p value using Benjamini Hochberg procedure)

| Structure | Mean CVN-AD (AU) | Mean Ctrl (AU) | Std AD (AU) | Std Ctrl (AU) | p | p BH | CI1 (AU) | CI2 (AU) | t | Cohen d | Difference (%) | Abbr | Hemisphere |
|---|---|---|---|---|---|---|---|---|---|---|---|---|---|
| Accumbens | 12505.47 | 12817.85 | 179.52 | 252.83 | 0.0024 | 0.0444 | -501.43 | -123.32 | -3.43 | -1.40 | -2.44 | Acb | Left |
| Caudomedial_Entorhinal_Cortex | 10441.80 | 10842.40 | 222.90 | 199.30 | 0.0001 | 0.0411 | -579.32 | -221.88 | -4.65 | -1.90 | -3.69 | CEnt | Right |
| Central_Gray | 11851.43 | 12530.45 | 524.97 | 392.64 | 0.0015 | 0.0444 | -1067.76 | -290.27 | -3.62 | -1.48 | -5.42 | CG | Left |
| Cingulate_Cortex_Area_24a | 12746.11 | 13240.90 | 349.27 | 270.58 | 0.0008 | 0.0430 | -757.19 | -232.40 | -3.91 | -1.60 | -3.74 | A24a | Right |
| Cingulate_Cortex_Area_24a_prime | 13218.88 | 13925.76 | 340.80 | 513.55 | 0.0008 | 0.0430 | -1083.65 | -330.12 | -3.89 | -1.59 | -5.08 | A24aPrime | Left |
| Cingulate_Cortex_Area_24a_prime | 12770.20 | 13512.56 | 276.76 | 580.38 | 0.0008 | 0.0430 | -1139.54 | -345.17 | -3.88 | -1.59 | -5.49 | A24aPrime | Right |
| Cingulate_Cortex_Area_24b | 13375.26 | 13790.90 | 260.77 | 325.50 | 0.0025 | 0.0444 | -668.68 | -162.61 | -3.41 | -1.40 | -3.01 | A24b | Left |
| Cingulate_Cortex_Area_29a | 12914.48 | 13558.58 | 332.18 | 515.22 | 0.0018 | 0.0444 | -1019.23 | -268.98 | -3.56 | -1.46 | -4.75 | A29a | Left |
| Cochlear_Nucleus | 10548.93 | 11258.91 | 511.75 | 448.31 | 0.0015 | 0.0444 | -1116.25 | -303.70 | -3.62 | -1.48 | -6.31 | CoN | Left |
| Facial_Nerve | 11288.89 | 11810.32 | 374.75 | 323.40 | 0.0014 | 0.0444 | -816.82 | -226.03 | -3.66 | -1.50 | -4.42 | n7 | Left |
| Fastigial_Medial_Dorsolateral_Nucleus_of_Cerebellum | 12012.35 | 12602.01 | 423.45 | 394.41 | 0.0019 | 0.0444 | -936.19 | -243.14 | -3.53 | -1.45 | -4.68 | FasDL | Left |
| Fornix | 12004.69 | 11714.81 | 207.74 | 204.16 | 0.0024 | 0.0444 | 115.03 | 464.73 | 3.44 | 1.41 | 2.47 | fx | Left |
| Hippocampus | 12742.53 | 13080.92 | 293.76 | 188.05 | 0.0025 | 0.0444 | -543.91 | -132.87 | -3.41 | -1.40 | -2.59 | Hc | Right |
| Inferior_Cerebellar_Peduncle | 8942.66 | 9442.07 | 397.22 | 221.96 | 0.0008 | 0.0430 | -766.18 | -232.63 | -3.88 | -1.59 | -5.29 | icp_oc_tz | Left |
| Middle_Cerebellar_Peduncle | 10080.60 | 10565.31 | 437.50 | 254.25 | 0.0027 | 0.0444 | -781.78 | -187.64 | -3.38 | -1.39 | -4.59 | mcp | Left |
| Parabrachial Koelliker_Fuse_Nucleus | 12237.93 | 12909.24 | 439.72 | 417.95 | 0.0009 | 0.0430 | -1034.93 | -307.69 | -3.83 | -1.57 | -5.20 | KF | Left |
| Parasubiculum | 12023.36 | 12546.10 | 478.83 | 271.04 | 0.0028 | 0.0448 | -845.47 | -200.02 | -3.36 | -1.38 | -4.17 | PaS | Right |
| Pedunculotegmental_Supratrigemnial_Nuclei | 12977.70 | 13590.20 | 510.47 | 370.33 | 0.0026 | 0.0444 | -985.99 | -239.01 | -3.40 | -1.39 | -4.51 | Su5 | Left |
| Primary_Somatosensory_Cortex_Jaw_Region | 12882.77 | 13186.41 | 183.84 | 227.39 | 0.0018 | 0.0444 | -480.97 | -126.30 | -3.55 | -1.45 | -2.30 | S1J | Right |
| Ventral_Spinocerebellar_Tract | 11587.32 | 12022.76 | 336.78 | 225.00 | 0.0010 | 0.0430 | -674.49 | -196.38 | -3.78 | -1.55 | -3.62 | vsc | Left |
| Vestibulocochlear_Nerve | 9880.86 | 10520.64 | 413.62 | 304.92 | 0.0003 | 0.0419 | -944.31 | -335.25 | -4.36 | -1.78 | -6.08 | n8 | Left |

**Supplementary Table 3. QSM differences in CVN-AD mice relative to *mNos2*-/- controls**
(Ctrl: *mNos2*-/-; Abbr: Abbreviation; p BH: corrected p value using Benjamini Hochberg procedure)

| Structure | Mean CVN-AD (ppm) | Mean Ctrl (ppm) | Std AD (ppm) | Std Ctrl (ppm) | p | p BH | CI1 (ppm) | CI2 (ppm) | t | Cohen d | Difference (%) | Abbr | Hemi-sphere |
|---|---|---|---|---|---|---|---|---|---|---|---|---|---|
| Fimbria | 0.0005 | -0.0034 | 0.0019 | 0.0020 | 0.0001 | 0.0054 | 0.0022 | 0.0055 | 4.82 | 1.97 | -113.73 | fi | Left |
| Hippocampus | 0.0000 | -0.0015 | 0.0006 | 0.0011 | 0.0009 | 0.0219 | 0.0007 | 0.0023 | 3.85 | 1.58 | -97.27 | Hc | Left |
| Insular_Cortex | -0.0006 | -0.0010 | 0.0003 | 0.0003 | 0.0026 | 0.0501 | 0.0002 | 0.0007 | 3.40 | 1.39 | -42.38 | Ins | Left |
| Interposed_Nucleus_of_Cerebellum | 0.0030 | 0.0077 | 0.0042 | 0.0024 | 0.0022 | 0.0466 | -0.0076 | -0.0019 | -3.47 | -1.42 | -61.70 | Int | Right |
| Lateral_Geniculate_Nucleus | 0.0013 | -0.0049 | 0.0025 | 0.0032 | 0.0000 | 0.0041 | 0.0037 | 0.0087 | 5.14 | 2.11 | -127.38 | LGN | Left |
| Medial_Orbital_Cortex | -0.0017 | -0.0059 | 0.0014 | 0.0020 | 0.0000 | 0.0022 | 0.0027 | 0.0057 | 5.87 | 2.41 | -71.20 | MO | Left |
| Medial_Orbital_Cortex | -0.0012 | -0.0044 | 0.0018 | 0.0022 | 0.0009 | 0.0219 | 0.0015 | 0.0050 | 3.85 | 1.58 | -72.49 | MO | Right |
| PosteriorDorsal_ParaventricularMedialParvicellular_Posterior_LateralHypothalamus | 0.0003 | 0.0031 | 0.0013 | 0.0016 | 0.0001 | 0.0068 | -0.0040 | -0.0015 | -4.59 | -1.88 | -89.10 | PHD_Lat Hy | Left |
| Pretectal_Nucleus | 0.0014 | -0.0059 | 0.0058 | 0.0035 | 0.0009 | 0.0219 | 0.0034 | 0.0114 | 3.82 | 1.57 | -124.15 | APT | Right |
| Primary_Somatosensory_Cortex | -0.0002 | -0.0027 | 0.0013 | 0.0016 | 0.0004 | 0.0169 | 0.0013 | 0.0037 | 4.16 | 1.70 | -92.57 | S1 | Right |
| Red_Nucleus_Magnocellular | -0.0041 | -0.0079 | 0.0023 | 0.0014 | 0.0000 | 0.0041 | 0.0022 | 0.0054 | 5.03 | 2.06 | -48.17 | RMC | Left |
| Ventral_Hippocampal_Commissure | 0.0003 | -0.0054 | 0.0043 | 0.0038 | 0.0023 | 0.0466 | 0.0023 | 0.0091 | 3.46 | 1.42 | -106.18 | vhc | Right |
| Ventral_Thalamic_Nuclei | -0.0003 | -0.0030 | 0.0019 | 0.0012 | 0.0005 | 0.0179 | 0.0013 | 0.0040 | 4.09 | 1.68 | -88.78 | VT | Right |
| Ventral_Thalamic_Nuclei | -0.0003 | -0.0025 | 0.0018 | 0.0009 | 0.0006 | 0.0206 | 0.0011 | 0.0034 | 3.99 | 1.63 | -87.13 | VT | Left |
| Periaquaductal_Grey | 0.0030 | 0.0002 | 0.0016 | 0.0019 | 0.0008 | 0.0219 | 0.0013 | 0.0043 | 3.87 | 1.58 | 1475.14 | PAG | Left |